\documentclass[12pt,a4paper]{article}

\usepackage[margin=1in,dvips,a4paper,nohead]{geometry}
\usepackage[margin=1em,font=small,justification=centerlast,tableposition=top]{caption}

\usepackage{pstricks}

\newdimen \psx
\newdimen \psy

\def\psddots (#1,#2){
  \psx #1\psxunit
  \psy #2\psyunit
  \qdisk(\psx,\psy){.03}
  \advance \psx -.2\psxunit
  \advance \psy .2\psyunit
  \qdisk(\psx,\psy){.03}
  \advance \psx .4\psxunit
  \advance \psy -.4\psyunit
  \qdisk(\psx,\psy){.03}
}

\def\pssddots (#1,#2){
  \psx #1\psxunit
  \psy #2\psyunit
  \qdisk(\psx,\psy){.03}
  \advance \psx .2\psxunit
  \advance \psy .2\psyunit
  \qdisk(\psx,\psy){.03}
  \advance \psx -.4\psxunit
  \advance \psy -.4\psyunit
  \qdisk(\psx,\psy){.03}
}

\usepackage[noBBpl,slantedGreek]{mathpazo}

\usepackage{amsfonts,amsmath,amssymb}
\allowdisplaybreaks

\usepackage{array}

\makeatletter
\def\eqnarray{%
   \stepcounter{equation}%
   \def\@currentlabel{\p@equation\theequation}%
   \global\@eqnswtrue
   \m@th
   \global\@eqcnt\z@
   \tabskip\@centering
   \let\\\@eqncr
   $$\everycr{}\halign to\displaywidth\bgroup
       \hskip\@centering$\displaystyle\tabskip\z@skip{##}$\@eqnsel
      &\global\@eqcnt\@ne\hfil$\displaystyle{\hbox{}##\hbox{}}$\hfil
      &\global\@eqcnt\tw@ $\displaystyle{##}$\hfil\tabskip\@centering
      &\global\@eqcnt\thr@@ \hb@xt@\z@\bgroup\hss##\egroup
         \tabskip\z@skip
      \cr
}
\DeclareRobustCommand
  \sddots{\mathinner{\mkern1mu\raise\p@\hbox{.}\mkern2mu
    \raise4\p@\hbox{.}\mkern2mu\raise7\p@
    \vbox{\kern7\p@\hbox{.}}\mkern1mu}}
\makeatother

\def\mbar#1{\kern 0.1em\overline{\kern -0.1em #1 \kern -0.1em}\kern
0.1em}

\def \Ad{\mathop{\mathrm{Ad}}}
\def \ad{\mathop{\mathrm{ad}}}
\def \Aut{\mathop{\mathrm{Aut}}}
\def \Der{\mathop{\mathrm{Der}}}
\def \id{\mathop{\mathrm{id}}}
\def \Inn{\mathop{\mathrm{Inn}}}
\def \Out{\mathop{\mathrm{Out}}}
\def \tr{\mathop{\mathrm{tr}}}

\def\bbR {\mathbb R}
\def\bbC {\mathbb C}
\def\bbZ {\mathbb Z}

\def\calF {\mathcal F}

\def\calL {\mathcal L}
\def\calM {\mathcal M}

\def\id{\mathop{\mathrm{id}}}

\def\gothd {\mathfrak d}
\def\gothg {\mathfrak g}
\def\gothgl {\mathfrak{gl}}
\def\gothh {\mathfrak h}
\def\gothso {\mathfrak{so}}
\def\gothsl {\mathfrak{sl}}
\def\gothsp {\mathfrak{sp}}

\def\rmC{\mathrm C}
\def\rmD {\mathrm D}

\def\rme {\mathrm e}
\def\rmi {\mathrm i}
\def\rmGL {\mathrm{GL}}
\def\rmO {\mathrm O}
\def\rmSL {\mathrm{SL}}
\def\rmSO {\mathrm{SO}}
\def\rmSp {\mathrm{Sp}}

\title{Toda equations associated with loop groups\\
of complex classical Lie groups}

\author{Kh. S. Nirov\\
\small \em Institute for Nuclear Research of the Russian Academy of
Sciences\\[-.3em]
\small \em 60th October Anniversary Prospect 7a, 117312 Moscow,
Russia\\[.3em]
A. V. Razumov\\
\small \em Institute for High Energy Physics\\[-.3em]
\small \em 142281 Protvino, Moscow Region, Russia}

\date{}

\begin{document}

\maketitle

\begin{abstract}
A Toda equation is specified by a choice of a Lie group and a $\mathbb
Z$-gradation of its Lie algebra. The Toda equations associated with
loop groups of complex classical Lie groups, whose Lie algebras are
endowed with integrable $\mathbb Z$-gradations with finite dimensional
grading subspaces, are described in an explicit form.
\end{abstract}

%\tableofcontents

\section{Introduction}

The Toda equations constitute a wide class of nonlinear integrable
equations arising in many mathematical and physical problems having as
fundamental as application significance.  Recall that according to the
general scheme, a concrete Toda equation arises when one chooses a Lie
group and specifies a $\bbZ$-gradation of its Lie algebra, see, for
example, the books \cite{LezSav92,RazSav97a}. Hence, to describe one
or another class of Toda equations one should start with some class of
$\bbZ$-gradations of Lie algebras. The Toda equations associated with
the complex classical Lie groups were classified and described in an
explicit form in the papers \cite{RazSav97b,RazSavZue99,NirRaz02}. In
the present paper we explicitly describe a wide class of Toda
equations associated with loop groups\footnote{In the present paper by
a loop group we mean either a usual loop group or a twisted loop
group. Similarly, by a loop Lie algebra we mean either a usual loop
Lie algebra or a twisted loop Lie algebra.} of classical Lie groups.

There are two main definitions of the loop Lie algebras. In accordance
with the first definition  a loop Lie algebra is the set of finite
Laurent polynomials with coefficients in a finite dimensional Lie
algebra, see, for example, the book \cite{Kac94}. The main
disadvantage of this definition is that it is impossible to associate
a Lie group with such a Lie algebra. However, using the results on
$\bbZ$-gradations of the affine Kac--Moody algebras \cite{Kac94}, in
the case when the underlying Lie algebra is complex and simple, one
can classify all $\bbZ$-gradations of the loop Lie algebras with
finite dimensional grading subspaces.

In accordance with the second definition a loop Lie algebra is the set
of smooth mappings from the circle $S^1$ to a finite dimensional Lie
algebra $\gothg$, see, for example, the book \cite{PreSea86}. This
definition is more convenient for applications to the theory of
integrable systems, because in this case we always have an appropriate
Lie group formed by smooth mappings from the circle $S^1$ to a Lie group
$G$ whose Lie algebra is $\gothg$. Therefore, it would be interesting and useful to obtain a
classification of $\bbZ$-gradations for loop Lie algebras defined as
in the book \cite{PreSea86}. This problem was partially solved in the paper
\cite{NirRaz06}. In that paper a concept of an integrable
$\bbZ$-gradation was introduced and all integrable $\bbZ$-gradations
with finite dimensional grading subspaces of loop Lie algebras of
finite dimensional complex simple Lie algebras were classified.
Any loop Lie algebra defined in accordance with the book
\cite{Kac94} is a subalgebra of the corresponding loop Lie algebra
defined in accordance with the book \cite{PreSea86}, and in the case
when the underlying Lie algebra is complex and simple any
$\bbZ$-gradation with finite dimensional grading subspaces of the
former is the restriction of an integrable $\bbZ$-gradation with finite
dimensional grading subspaces of the latter.

In the
present paper we use the results of the paper \cite{NirRaz06} to
describe in an explicit form the Toda equations associated with loop
groups of complex classical Lie groups based on integrable
$\bbZ$-gradations with finite dimensional grading subspaces. Here we
use the convenient block matrix representations for the elements of
the corresponding classical Lie algebras. This representation appeared
to be effective to analyze the Toda equations associated with finite
dimensional Lie groups
\cite{RazSav97b, RazSavZue99, NirRaz02, NirRaz02a, NirRaz02b}.

There is a large number of works devoted to Toda equations associated
with loop groups. From our point of view, an incomplete list of references
which are the most relevant to the contents of the present investigation
consists of the papers \cite{Mik81, MikOlsPer81, LezSav83,
DelFeh95, FerMirGui95, FerMirGui97, FerGalHolMir97, Mir99}.

There exist so-called higher grading
\cite{Lez85,GerSav95,FerGerSanSav96,BueFerRaz02} and multi-dimensional
\cite{RazSav97c, RazSav97d} generalizations of the Toda equations. The
approach of the present paper can be generalized and used for
classification of those equations.

Concluding this introductory section let us describe the notation used
in the paper. We denote by $I_n$ the unit diagonal $n \times n$ matrix
and by $J_n$ the unit skew diagonal $n \times n$ matrix. For an even
$n$ we also define
\[
K_n = \left( \begin{array}{cc}
0 & J_{n/2} \\
- J_{n/2} & 0
\end{array} \right).
\]
When it does not lead to a misunderstanding, we write instead of
$I_n$, $J_n$, and $K_n$ just $I$, $J$, and $K$ respectively.

If $m$ is an $n_1 \times n_2$ matrix, $A$ is an $n_2 \times n_2$
nonsingular matrix, $B$ is an $n_1 \times n_1$ matrix, we denote
\[
{}^{AB\!} m = A^{-1} \, {}^{t\!} m \, B,
\]
where ${}^{t\!} m$ is the transpose of the matrix $m$. We also write
${}^A m$ instead of ${}^{AA} m$. Note that ${}^{J\!} m$ is actually
the transpose of $m$ with respect to the skew diagonal.

It is useful to have in mind that ${}^A({}^{A\!} m) = B^{-1} m \, B$,
where $B = {}^{t\!}(A^{-1}) A$. In particular, one can show that
${}^J({}^{J\!} m) = m$ and ${}^K({}^{K\!} m) = m$.

\section{Toda equations associated with loop groups}

\subsection{Loop Lie algebras and loop groups}

We start this section with an explanation of what we mean by loop Lie
algebras and loop groups. Here we mainly follow the book
\cite{PreSea86} and the papers \cite{Ham82,Mil84,NirRaz06}.

Let $\gothg$ be a real or complex Lie algebra. Define the {\em loop
Lie algebra\/} $\calL(\gothg)$ as the linear space $C^\infty(S^1,
\gothg)$ of smooth mappings from the circle $S^1$ to $\gothg$ with the
Lie algebra operation defined pointwise. We think of the circle $S^1$
as consisting of complex numbers of modulus one. Since $\calL(\gothg)$
is infinite dimensional, one has to deal with infinite sums of its
elements. Therefore, it is necessary to have some topology on
$\calL(\gothg)$ consistent with the Lie algebra operation. We assume
that the required topology is introduced by supplying $\calL(\gothg)$
in an appropriate way with the structure of a Fr\'echet space, see,
for example, \cite{Ham82,Mil84,NirRaz06}.

Now, let $G$ be a Lie group with the Lie algebra $\gothg$. The {\rm
loop group\/} $\calL(G)$ is the set $C^\infty(S^1, G)$ with the group
law defined pointwise. We endow $\calL(G)$ with the structure of a
Fr\'echet manifold modeled on $\calL(\gothg)$ in such a way that it
becomes a Lie group, see, for example, \cite{Ham82,Mil84,NirRaz06}.
Here the Lie algebra of the Lie group $\calL(G)$ is naturally
identified with the loop Lie algebra~$\calL(\gothg)$.

Let $A$ be an automorphism of a Lie algebra $\gothg$ satisfying the
relation $A^M = \id\nolimits_\gothg$ for some positive integer $M$.
The {\em twisted loop Lie algebra\/} $\calL_{A,M}(\gothg)$ is a
subalgebra of the loop Lie algebra $\calL(\gothg)$ formed by the
elements $\xi$ which satisfy the equality
\[
\xi(\epsilon_M p) = A(\xi(p)),
\]
where $\epsilon_M = \exp(2 \pi \rmi/M)$ is the $M$th principal root of
unity. Similarly, given an automorphism $a$ of a Lie group $G$ which
satisfies the relation $a^M = \id_G$, we define the {\em twisted loop
group\/} $\calL_{a,M}(G)$ as the subgroup of the loop group $\calL(G)$
formed by the elements $\gamma$ satisfying the equality
\[
\gamma(\epsilon_M p) = a(\gamma(p)).
\]
The Lie algebra of a twisted loop group $\calL_{a,M}(G)$ is naturally
identified with the twis\-ted loop Lie algebra $\calL_{A,M}(\gothg)$,
where we denote the automorphism of the Lie algebra $\gothg$
corresponding to the automorphism $a$ of the Lie group $G$ by $A$.

It is clear that a loop Lie algebra $\calL(\gothg)$ can be treated as
a twisted loop Lie algebra $\calL_{\id_\gothg, M}(\gothg)$, where $M$
is an arbitrary positive integer. In its turn, a loop group $\calL(G)$
can be treated as a twisted loop group $\calL_{\id_G, M}(G)$, where
$M$ is again an arbitrary positive integer. In the present paper by
loop Lie algebras and loop groups we mean twisted loop Lie algebras
and twisted loop groups.

To construct a Toda equation associated with a Lie group one should
first endow its Lie algebra with a $\bbZ$-gradation. We say that the
loop Lie algebra $\calL_{A,M}(\gothg)$ is endowed with a {\em
$\bbZ$-gradation\/} if for any $k \in \bbZ$ there is given a closed
subspace $\calL_{A, M}(\gothg)_k$ of $\calL_{A, M}(\gothg)$ such that

(a) for any $k, l \in \bbZ$ one has $[\calL_{A, M}(\gothg)_k,
\calL_{A, M}(\gothg)_l] \subset \calL_{A,M}(\gothg)_{k+l}$,

(b) any element $\xi$ of $\calL_{A,M}(\gothg)$ can be uniquely
represented as an absolutely convergent series
\[
\xi = \sum_{k \in \bbZ} \xi_k,
\]
where $\xi_k \in \calL_{A, M}(\gothg)_k$. The subspaces $\calL_{A,
M}(\gothg)_k$ are called the {\em grading subspaces\/} of $\calL_{A,
M}(\gothg)$ and the elements $\xi_k$ the {\em grading components\/} of
$\xi$. Note that $\calL_{A, M}(\gothg)_0$ is a subalgebra of
$\calL_{A, M}(\gothg)$.

Let $\gothg$ and $\gothh$ be two Lie algebras, and $F$ be an
isomorphism from a loop Lie algebra $\calL_{A, M}(\gothg)$ to a loop
Lie algebra $\calL_{B, N}(\gothh)$. Taking the subspaces $\calL_{B,
N}(\gothh)_k = F(\calL_{A, M}(\gothg)_k)$ as grading subspaces we
endow $\calL_{B,N}(\gothh)$ with a $\bbZ$-gradation.

In general, if the grading subspaces of two loop Lie algebras
$\calL_{A, M}(\gothg)$ and $\calL_{B, N}(\gothh)$ are connected by the
relation $\calL_{B, N}(\gothh)_k = F(\calL_{A, M}(\gothg)_k)$, where
$F$ is an isomorphism from $\calL_{A, M}(\gothg)$ to $\calL_{B,
N}(\gothh)$, we say that the corresponding $\bbZ$-gradations are {\em
conjugated by\/} the {\em isomorphism\/} $F$. It is clear that if the
grading components of an element $\xi \in \calL_{A, M}(\gothg)$ are
$\xi_k$, then the grading components of the element $F(\xi) \in
\calL_{B,N}(\gothh)$ are $F(\xi_k)$.

\subsection{Toda equations}

Let $\calM$ be either the real manifold $\bbR^2$ or the complex
manifold $\mathbb C$. Denote the standard coordinates on $\bbR^2$ by
$z^-$ and $z^+$. In the case of the manifold $\bbC$ we denote by $z^-$
the standard complex coordinate $z$ and by $z^+$ its complex conjugate
$\bar z$.\footnote{Actually one can assume that $\calM$ is an
arbitrary two-dimensional real manifold, or one-dimensional complex
manifold, see, for example, the paper \cite{RazSav94} and the book
\cite{RazSav97a}. Here $z^-$ and $z^+$ are some local coordinates.} We
use the usual notation
\[
\partial_- = \partial/\partial z^-, \quad \partial_+ =
\partial/\partial z^
+.
\]

Let $G$ be a complex simple Lie group. The Lie algebra $\gothg$ of $G$
is certainly simple. Actually without any loss of generality one can
assume that $G$ is a matrix Lie group. Consider a loop group
$\calL_{a, M}(G)$ and the corresponding loop Lie algebra $\calL_{A,
M}(\gothg)$. Suppose that $\calL_{A, M}(\gothg)$ is endowed with a
$\bbZ$-gradation. Let for some positive integer $L$ the subspaces
$\calL_{A, M}(\gothg)_{-k}$ and $\calL_{A, M}(\gothg)_{+k}$ for $0 < k
< L$ be trivial. Denote by $\calL_{a, M}(G)_0$ the connected Lie
subgroup of $\calL_{a,M}(G)$ corresponding to the subalgebra
$\calL_{A, M}(\gothg)_0$. The {\em Toda equation\/} associated with
the loop group $\calL_{a, M}(G)$ is an equation for a smooth mapping
$\Xi$ from $\calM$ to $\calL_{a, M}(G)_0$ which has the following
form
\begin{equation}
\partial_+(\Xi^{-1} \partial_- \Xi) = [\calF_-, \Xi^{-1}
\calF_+ \Xi], \label{e:1}
\end{equation}
see, for example, the books \cite{LezSav92,RazSav97a}. Here $\calF_-$
and $\calF_+$ are some fixed smooth mappings from $\calM$ to
$\calL_{A, M}(\gothg)_{-L}$ and $\calL_{A, M}(\gothg)_{+L}$,
respectively, satisfying the conditions
\begin{equation}
\partial_+ \calF_- = 0, \qquad \partial_- \calF_+ = 0. \label{e:2}
\end{equation}
When the Lie group $\calL_{a, M}(G)_0$ is Abelian, the corresponding
Toda equation is said to be {\em Abelian\/}, otherwise we deal with a
{\em non-Abelian Toda equation\/}.

The authors of the paper \cite{FerMirGui95} consider equations of the
form (\ref{e:1}) for the case when $L$ does not satisfy the condition
that the subspaces $\calL_{A, M}(\gothg)_{-k}$ and $\calL_{A,
M}(\gothg)_{+k}$ for $0 < k < L$ are trivial. Actually it is convenient
to consider such equations as a reduction of higher grading Toda
equations \cite{Lez85,GerSav95,FerGerSanSav96}.

It is clear that to classify Toda equations associated with the loop
groups $\calL_{a, M}(G)$ one has to classify $\bbZ$-gradations of the
loop Lie algebras $\calL_{A, M}(\gothg)$. Here, if we endow the Lie
algebras of two loop groups connected by an isomorphism by conjugated
$\bbZ$-gradations, we will obtain equivalent Toda equations.

In the present paper we restrict ourselves to the case of integrable
$\bbZ$-gradations with finite dimensional grading subspaces. In
accordance with the definition given in the paper~\cite{NirRaz06} a
$\bbZ$-gradation of $\calL_{A, M}(\gothg)$ is {\em integrable\/} if
the mapping
\[
(\tau, \xi) \in \bbR \times \calL_{A, M}(\gothg) \mapsto \sum_{k \in
\bbZ} \rme^{- \rmi k \tau} \xi_k \in \calL_{A, M}(\gothg)
\]
is smooth. As usually, we denote by $\xi_k$ the grading components of
the element $\xi$ with respect to the $\bbZ$-gradation under
consideration.

An important example of an integrable $\bbZ$-gradation is the {\em
standard gradation\/} of a loop Lie algebra $\calL_{A, M}(\gothg)$.
Here the grading subspaces are
\[
\calL_{A, M}(\gothg)_{k} = \{ \xi = \lambda^k x \in
\calL_{A,M}(\gothg) \mid x \in \gothg, \ A(x) = \epsilon_M^k x \},
\]
where by $\lambda$ we denote the restriction of the standard
coordinate on $\bbC$ to $S^1$.

Suppose that the loop Lie algebra $\calL_{A, M}(\gothg)$ is endowed
with an integrable $\bbZ$-gradation with finite dimensional grading
subspaces. As was proved in the paper~\cite{NirRaz06}, this
$\bbZ$-gradation is conjugated by an isomorphism to the standard
$\bbZ$-gradation of some other loop Lie algebra $\calL_{B,
N}(\gothg)$. Here the automorphisms $A$ and $B$ differ by an inner
automorphism of $\gothg$. For the case of Toda equations the
automorphism $A$ of the Lie algebra $\gothg$ should be generated by an
automorphism $a$ of the Lie group $G$. It is clear that in this case
the automorphism $B$ can be lifted up to an automorphism $b$ of $G$.
Moreover, one can show that the isomorphism that conjugates the
$\bbZ$-gradation of the loop Lie algebra $\calL_{A, M}(\gothg)$ with
the standard $\bbZ$-gradation of the loop Lie algebra $\calL_{B,
N}(\gothg)$ can be lifted to an isomorphism connecting the loop groups
$\calL_{a, M}(G)$ and $\calL_{b, N}(G)$. Thus, the corresponding Toda
equations associated with the loop groups $\calL_{a, M}(G)$ and
$\calL_{b, N}(G)$ are equivalent. Therefore, to describe the Toda
equations belonging to the class under consideration it suffices to
classify the Lie groups $\calL_{a, M}(G)$. This problem is evidently
equivalent to classification of finite order automorphisms of $G$.

Let $c$ be an arbitrary automorphism of $G$. Note that the mapping
$\gamma  \mapsto c \circ \gamma$ is an isomorphism from a loop group
$\calL_{a, M}(G)$ to the loop group $\calL_{b, M}(G)$, where $b = c
\circ a \circ c^{-1}$. This isomorphism generates the isomorphism from
the loop Lie algebra $\calL_{A, M}(\gothg)$ to the loop Lie algebra
$\calL_{B, M}(\gothg)$, where $B$ is the automorphism of $\gothg$
corresponding to the automorphism $b$ of $G$.  It is clear that the
standard gradations of $\calL_{A, M}(\gothg)$ and $\calL_{B,
M}(\gothg)$ are conjugated by the isomorphism under consideration, and
the corresponding Toda equations are equivalent. Therefore, to
classify Toda equations it suffices to classify finite order
automorphisms of $G$ up to conjugation.

Note that any automorphism $A$ of $\gothg$ satisfying the relation
$A^M = \id_{\gothg}$ determines a $\bbZ_M$-gradation of $\gothg$.
Indeed, since the automorphism $A$ is of finite order, it is
semisimple. The eigenvalues of $A$ are of the form $\epsilon_M^k$,
where $0 \le k \le M-1$. Denote by $[k]_M$ the element of the ring
$\bbZ_M$ corresponding to the integer $k$, and define the subspaces
$\gothg_{[k]_M}$ of $\gothg$ as
\[
\gothg_{[k]_M} = \{x \in \gothg \mid  A(x) = \epsilon_M^k x\}.
\]
It is evident that
\[
\gothg = \bigoplus_{k=0}^{M-1} \gothg_{[k]_M} = \bigoplus_{s \in
\bbZ_M} \gothg_s
\]
and this decomposition is a $\bbZ_M$-gradation of $\gothg$. Vice
versa, any $\bbZ_M$-gradation of $\gothg$ determines in an evident way
an automorphism $A$ of $\gothg$ which satisfies the relation $A^M =
\id_{\gothg}$. If $A$ is an inner automorphism of $\gothg$ we say that
the corresponding $\bbZ_M$-gradation is of {\em inner type\/},
otherwise we say that it is of {\em outer type\/}. In terms of the
$\bbZ_M$-gradation corresponding to the automorphism $A$, the grading
subspaces of the standard $\bbZ$-gradation of the loop Lie algebra
$\calL_{A, M}(\gothg)$ can be described as
\[
\calL_{A, M}(\gothg)_k = \{ \xi \in \calL_{A, M}(\gothg) \mid \xi =
\lambda^k x, \ x \in \gothg_{[k]_M} \}.
\]
Above we denoted by $L$ a positive integer such that the grading
subspaces with the grading index $k$ satisfying the condition $0 < |k|
< L$ are trivial. It is not difficult to understand that for the
standard $\bbZ$-gradation $L \le M$. The case $L = M$ arises if and
only if $A = \id_{\gothg}$ and $M$ is an arbitrary positive integer.
In this case the nontrivial grading subspaces are $\calL_{\id_\gothg,
M}(\gothg)_{kM}$, $k \in \bbZ$, and it is clear that
\[
\calL_{\id_\gothg, M}(\gothg)_{kM} = \{ \xi \in \calL_{\id_\gothg, M}(\gothg)
\mid \xi = \lambda^{kM} x, \ x \in \gothg \}.
\]

It is clear that for the standard $\bbZ$-gradation the subalgebra
$\calL_{A, M}(\gothg)_0$ is isomorphic to $\gothg_{[0]_M}$, and the
Lie group $\calL_{a, M}(G)_0$ is isomorphic to the connected Lie
subgroup $G_0$ of $G$ corresponding to the Lie algebra
$\gothg_{[0]_M}$. Hence, the mapping $\Xi$ is actually a mapping
from $\calM$ to $G_0$, for consistency with the notation used earlier
we will denote it by $\gamma$. The mappings $\calF_-$ and $\calF_+$
are given by the relation
\[
\calF_-(p) = \lambda^{-L} c_-(p), \qquad \calF_+(p) = \lambda^L c_+(p),
\qquad p \in \calM,
\]
where $c_-$ and $c_+$ are
mappings from $\calM$ to $\gothg_{-[L]_M}$ and $\gothg_{+[L]_M}$
respectively. Thus, the Toda equation (\ref{e:1}) can be written as
\begin{equation}
\partial_+(\gamma^{-1} \partial_- \gamma) = [c_-, \gamma^{-1} c_+
\gamma], \label{e:5}
\end{equation}
where $\gamma$ is a smooth mapping from $\calM$ to the connected Lie
subgroup $G_0$ of $G$ corresponding to the Lie algebra
$\gothg_{[0]_M}$, and the mappings $c_-$ and $c_+$ are fixed smooth
mappings from $\calM$ to $\gothg_{-[L]_M}$ and $\gothg_{+[L]_M}$
respectively. The conditions (\ref{e:2}) imply that
\begin{equation}
\partial_+ c_- = 0, \qquad \partial_- c_+ = 0. \label{e:6}
\end{equation}
Summarizing one can say that a Toda equation associated with a loop
group of a simple complex Lie group whose Lie algebra is endowed with
an integrable $\bbZ$-gradation with finite dimensional grading subspaces
is equivalent to the equation of the form~(\ref{e:5}).

The simplest case is when $a = \id_G$, $A = \id_\gothg$ and $M$ is an arbitrary
positive number. Remind that in this case $L = M$. The mapping
$\gamma$ is a mapping from $\calM$ to $G$, and the mappings $c_+$ and
$c_-$ are mappings from $\calM$ to $\gothg$. For consistency with the
notation used below we redenote $\gamma$ by $\Gamma$, $c_+$ by $C_+$,
and $c_-$ by $C_-$. Here the Toda equation (\ref{e:5}) becomes
\begin{equation}
\partial_+(\Gamma^{-1} \partial_- \Gamma) = [C_-, \Gamma^{-1} C_+
\Gamma], \label{e:5a}
\end{equation}
and the conditions (\ref{e:6}) read
\begin{equation}
\partial_+ C_- = 0, \qquad \partial_- C_+ = 0. \label{e:6a}
\end{equation}

It is natural to consider an equation of the type (\ref{e:5}) in a
more general setting. Namely, let $G$ be an arbitrary finite
dimensional Lie group and $a$ be an arbitrary finite order
automorphism of $G$. The corresponding automorphism $A$ of the Lie
algebra $\gothg$ of the Lie group $G$ generates a $\bbZ_M$-gradation
of $\gothg$. Assume that for some positive integer $L \le M$ the
grading subspaces $\gothg_{+[k]_M}$ and $\gothg_{-[k]_M}$ for $0 < k <
L$ are trivial. Choose some fixed mappings $c_+$ and $c_-$ from
$\calM$ to $\gothg_{+[L]_M}$ and $\gothg_{-[L]_M}$, respectively,
satisfying the relations (\ref{e:6}). Now Eq.~(\ref{e:5}) is
equivalent to the Toda equation associated with the loop group
$\calL_{a, M}(G)$ whose Lie algebra $\calL_{A, M}(\gothg)$ is endowed
with the standard $\bbZ$-gradation.

Note that the authors of the paper
\cite{FerGalHolMir97} also suggest Eq.~(\ref{e:5}) as a
convenient form of a Toda equation associated with a loop group.
They also explicitly show how to go back from Eq.~(\ref{e:5})
to a Toda equation associated with a loop group. Repeat that we proved
that in the case when $G$ is a simple complex Lie group such a procedure
gives all nonequivalent Toda equations for $\calL_{a,M}(G)$ based on
integrable $\bbZ$-gradations with finite dimensional grading subspaces.

Below we
describe a concrete form which Eq.~(\ref{e:5}) takes in the
case when $G$ is a complex classical Lie group. To this end we
classify up to conjugation the finite order automorphisms of the
corresponding Lie algebras. Such a classification was performed
earlier using root technique, see, for example, \cite{Kac94, GorOniVin94}.
Although this approach gives an answer, it appears to be inconvenient
for description of the structure of grading subspaces that is important
for us. Therefore, we develop and use another classification
based on the appropriate block matrix representations of the Lie
algebras under consideration.

\section{Toda equations associated with loop groups of complex general
linear groups. Gradations of inner type}

Let $a$ be an inner automorphism of the Lie group $\rmGL_n(\bbC)$
satisfying the relation $a^M = \id_{\rmGL_n(\bbC)}$. Denote the
corresponding inner automorphism of the Lie algebra $\gothgl_n(\bbC)$
by $A$. This automorphism satisfies the relation $A^M =
\id_{\gothgl_n(\bbC)}$. In other words, $A$ is a finite order
automorphism of $\gothgl_n(\bbC)$. Since we are interested in the
automorphisms of $\gothgl_n(\bbC)$ up to conjugations, we assume that
the automorphism $A$ under consideration is given by the relation
\[
A(x) = h \, x \, h^{-1},
\]
where $h$ is an element of the subgroup $\rmD_n(\bbC)$ of
$\rmGL_n(\bbC)$ formed by all complex nonsingular diagonal matrices,
see Appendix \ref{a:c2}. It is clear that multiplying $h$ by an
arbitrary nonzero complex number we obtain an element of
$\rmD_n(\bbC)$ which generates the same automorphism of
$\gothgl_n(\mathbb C)$ as the initial element.

The equality $A^M = \id_{\gothgl_n(\bbC)}$ gives $h^M \, x \, h^{-M} =
x$ for any $x \in \gothgl_n(\mathbb C)$. Therefore, $h^M = \nu I$,
where $\nu$ is a nonzero complex number.

Using inner automorphisms of $\gothgl_n(\mathbb C)$ which permute the
rows and columns of the matrix $h$ synchronously, we collect
coinciding diagonal matrix elements together, and come to the
following block diagonal form of the element $h$:
\begin{equation}
h = \left( \begin{array}{cccc}
\mu_1 I_{n_1} & & & \\
& \mu_2 I_{n_2} & & \\
& & \ddots & \\
& & & \mu_p I_{n_p}
\end{array} \right). \label{e:7}
\end{equation}
Here $n_\alpha$, $\alpha = 1, \ldots, p$, are positive integers, such
that $\sum_{\alpha=1}^p n_\alpha = n$, and $\mu_\alpha$ are complex
numbers, such that $\mu_\alpha^M = \nu$ for each $\alpha = 1, \ldots,
p$. We assume that $p > 1$. The case $p = 1$ corresponds to $A =
\id_{\gothgl_n(\bbC)}$. Here we come to Eq.~(\ref{e:5a}),
where $\Gamma$ is a mapping from $\calM$ to $\rmGL_n(\bbC)$, $C_+$ and
$C_-$ are mappings from $\calM$ to $\gothgl_n(\bbC)$ satisfying the
conditions (\ref{e:6a}).

Let $\rho$ be any complex number such that $\rho^M = \nu$. The number
$\rho$ is defined up to multiplication by an $M$th root of unity.
Represent the numbers $\mu_\alpha$ in the form
\begin{equation}
\mu_\alpha = \rho \, \epsilon_M^{m_\alpha}, \label{e:8}
\end{equation}
where $m_\alpha$ are integers satisfying the condition $0 < m_\alpha
\le M$. It is clear that $p \le M$. Without loss of generality, we
assume that $m_1 > m_2 > \ldots > m_p$. This can be provided by an
appropriate inner automorphism of $\gothgl_n(\mathbb C)$. Thus, we
assume that $h$ has the form
\begin{equation}
h = \rho \left( \begin{array}{cccc}
\epsilon_M^{m_1} I_{n_1} & & & \\
& \epsilon_M^{m_2} I_{n_2} & & \\
& & \ddots & \\
& & & \epsilon_M^{m_p} I_{n_p}
\end{array} \right), \label{e:9}
\end{equation}
where $M \ge m_1 > m_2 > \ldots > m_p> 0$.

Now consider the corresponding $\bbZ$-gradation. Represent a general
element $x$ of $\gothgl_n(\mathbb C)$ in the block matrix form
\begin{equation}
x = \left( \begin{array}{cccc}
x_{11} & x_{12} & \cdots & x_{1p} \\[.5em]
x_{21} & x_{22} & \cdots & x_{2p} \\[.5em]
\vdots & \vdots & \ddots & \vdots \\[.5em]
x_{p1} & x_{p2} & \cdots & x_{pp}
\end{array} \right), \label{e:10}
\end{equation}
where $x_{\alpha \beta}$ is an $n_\alpha \times n_\beta$ matrix. It is
clear that
\[
(h \, x \, h^{-1})_{\alpha \beta}^{} = \mu_\alpha^{} \mu_\beta^{-1}
x_{\alpha
\beta}^{} = \epsilon_M^{m_\alpha - m_\beta} x_{\alpha \beta}.
\]
Hence, if for fixed $\alpha$ and $\beta$ only the block $x_{\alpha
\beta}$ of the element $x$ is different from zero, then $x$ belongs to
the grading subspace $[m_\alpha - m_\beta]_M$. It is convenient to
introduce integers $k_\alpha$, $\alpha = 1, \ldots, p-1$, defined as
$k_\alpha = m_\alpha - m_{\alpha + 1}$. By definition, for each
$\alpha$ the integer $k_\alpha$ is positive and $\sum_{\alpha=1}^{p-1}
k_\alpha = m_1 - m_p < M$. It is clear that for $\alpha < \beta$ one
has
\[
[m_\alpha - m_\beta]_M = [\sum_{\gamma=\alpha}^{\beta-1} k_\gamma]_M,
\]
and for $\alpha > \beta$
\[
[m_\alpha - m_\beta]_M = -[m_\beta - m_\alpha]_M = -
[\sum_{\gamma=\beta}^{\alpha-1} k_\gamma]_M = [M -
\sum_{\gamma=\beta}^{\alpha-1} k_\gamma]_M.
\]
Now one can easily understand that the grading structure of the
$\bbZ_M$-gradation generated by the automorphism $A$ under
consideration can be depicted by the scheme given in Figure \ref{f:1}.
\begin{figure}[htb]
\[
\left( \begin{array}{c|c|c|c|c}
[0]_M & [k_1]_M & [k_1 + k_2]_M & \cdots & \displaystyle
[\sum_{\alpha=1}^{p-1} k_\alpha]_M \\ \hline
-[k_1]_M & [0]_M & [k_2]_M & \cdots & \displaystyle
[\sum_{\alpha=2}^{p-1}
k_\alpha]_M \\ \hline
-[k_1 + k_2]_M & - [k_2]_M & [0]_M & \cdots & \displaystyle
[\sum_{\alpha = 3}^{p-1} k_\alpha]_M \\ \hline
\vdots & \vdots & \vdots & \ddots & \vdots \\ \hline
\displaystyle -[\sum_{\alpha = 1}^{p-1} k_\alpha]_M & \displaystyle
- [\sum_{\alpha=2}^{p-1} k_\alpha]_M & \displaystyle -
[\sum_{\alpha=3}^{p-1}
k_\alpha]_M & \cdots & [0]_M
\end{array} \right)
\]
\caption{The canonical structure of a $\bbZ_M$-gradation} \label{f:1}
\end{figure}
Here the elements of the ring $\bbZ_M$ are the grading indices of the
corresponding blocks in the block matrix representation (\ref{e:10})
of a general element of $\mathfrak{gl}_n(\mathbb C)$. Note, in
particular, that the subalgebra $\gothg_{[0]_M}$ is formed by all
block diagonal matrices and is isomorphic to the Lie algebra
$\gothgl_{n_1}(\mathbb C) \times \cdots \times \gothgl_{n_p}(\mathbb
C)$. The group $G_0$ is also formed by block diagonal matrices and is
isomorphic to $\mathrm{GL}_{n_1}(\mathbb C) \times \cdots \times
\mathrm{GL}_{n_p}(\mathbb C)$.

Assume now that there are given $p$ positive integers $n_\alpha$ such
that $\sum_{\alpha=1}^p n_\alpha = n$ and $p-1$ positive integers
$k_\alpha$ such that $\sum_{\alpha=1}^{p-1} k_\alpha < M$. Let $m_p$
be a positive integer such that $\sum_{\alpha=1}^{p-1} k_\alpha +
m_p \le M$. Using the relation
\[
m_\alpha = \sum_{\beta=\alpha}^{p-1} k_\beta + m_p, \qquad \alpha = 1,
\ldots, p-1,
\]
one defines the remaining $p-1$ numbers $m_\alpha$. Now using
(\ref{e:8}) with an arbitrary nonzero $\rho$, we obtain the numbers
$\mu_\alpha$ and construct an element $h \in \rmGL_n(\bbC)$ in
accordance with (\ref{e:7}). This element determines an inner
automorphism of $\gothgl_n(\bbC)$ of order $M$ which in its turn
generates a $\bbZ_M$-gradation of $\gothgl_n(\bbC)$ corresponding to
our choice of numbers $n_\alpha$ and $k_\alpha$. Note that using the
arbitrariness of the number $\rho$ one can make $h$ meet some
additional requirements. In particular, one can always choose $h$ so
that $\det h = 1$.

Thus, up to a conjugation, a $\bbZ_M$-gradation of $\gothgl_n(\bbC)$
of inner type can be specified by a choice of $p \le n$ positive
integers $n_\alpha$, satisfying the equality $\sum_{\alpha=1}^p
n_\alpha = n$, and $p-1$ positive integers $k_\alpha$, satisfying the
inequality $\sum_{\alpha = 1}^{p-1} k_\alpha < M$.

Below we show that a $\bbZ_M$-gradation of an arbitrary complex
classical Lie algebra is characterized by $p$ positive integers
$n_\alpha$ and $p-1$ positive integers $k_\alpha$, and up to the
conjugation by an isomorphism the structure of the $\bbZ_M$-gradation
is described by Figure \ref{f:1}, except for the case considered in
Section \ref{s:6}. Therefore, we call this structure canonical. The
only difference from the case of complex general linear groups is that
for other groups the numbers $n_\alpha$, $k_\alpha$ and the blocks of
the block matrix representation (\ref{e:10}) satisfy some additional
conditions.

Consider now the corresponding Toda equations. Choose some
$\bbZ_M$-gradation of inner type of the Lie algebra $\gothg =
\gothgl_n(\bbC)$. Let $L$ be a positive integer such that the grading
subspaces $\gothg_{+[k]_M}$ and $\gothg_{-[k]_M}$ for $0 < k < L$ are
trivial. One can get convinced that if $x \in \gothg_{+[L]_M}$, then
only the blocks $x_{\alpha, \alpha+1}$, $\alpha = 1, \ldots, p-1$, and
$x_{p1}$ in the block matrix representation (\ref{e:10}) can be
different from zero. Thus, the mapping $c_+$ has the structure given
in Figure \ref{f:2},
\begin{figure}[ht]
\[
\psset{xunit=2.7em, yunit=1.6em}
\left( \begin{pspicture}[.5](.4,.5)(5.6,5.3)
\rput(1,5){$0$} \rput(2,4.92){$C_{+1}$}
\rput(2,4){$0$} \psddots(3,4)
\psddots(3,3) \psddots(4,3)
\rput(4,2){$0$} \rput(5,1.87){$C_{+(p-1)}$}
\rput(1,.94){$C_{+0}$} \rput(5,1){$0$}
\end{pspicture} \right)
\]
\caption{The canonical structure of the mapping  $c_+$} \label{f:2}
\end{figure}
where for each $\alpha = 1, \ldots, p-1$ the mapping $C_{+\alpha}$ is
a mapping from $\calM$ to the space of $n_\alpha \times n_{\alpha+1}$
complex matrices, and $C_{+0}$ is a mapping from $\calM$ to the space
of $n_p \times n_1$ complex matrices. Here we assume that if some
blocks among $x_{\alpha, \alpha+1}$, $\alpha = 1, \ldots, p-1$, and
$x_{p1}$ in the general block matrix representation (\ref{e:10}) have
the grading index different from $+[L]_M$, then the corresponding
blocks in the block representation of $c_+$ are zero matrices.

Similarly, we see that the mapping $c_-$ has the structure given in
Figure \ref{f:3},
\begin{figure}[ht]
\[
\psset{xunit=2.7em, yunit=1.6em}
\left( \begin{pspicture}[.5](.4,.5)(5.6,5.3)
\rput(1,5){$0$} \rput(5,4.92){$C_{-0}$}
\rput(1,4){$C_{-1}$} \rput(2,4){$0$}
\psddots(2,3) \psddots(3,3)
\psddots(3,2) \rput(4,1.87){$0$}
\rput(4,.94){$C_{-(p-1)}$} \rput(5,1){$0$}
\end{pspicture} \right)
\]
\caption{The canonical structure of the mapping  $c_-$} \label{f:3}
\end{figure}
where for each $\alpha = 1, \ldots, p-1$ the mapping $C_{-\alpha}$ is
a mapping from $\calM$ to the space of $n_{\alpha+1} \times n_\alpha$
complex matrices, and $C_{-0}$ is a mapping from $\calM$ to the space
of $n_1 \times n_p$ complex matrices. Here we assume that if some
blocks among $x_{\alpha+1, \alpha}$, $\alpha = 1, \ldots, p-1$, and
$x_{1p}$ in the general block matrix representation (\ref{e:10}) have
the grading index different from $-[L]_M$, then the corresponding
blocks in the block representation of $c_-$ are zero matrices. It is
clear that to provide the validity of the conditions (\ref{e:6}) the
mappings $C_{-\alpha}$ and $C_{+\alpha}$, $\alpha = 0, \ldots, p-1$,
must satisfy the relations
\begin{equation}
\partial_+ C_{-\alpha} = 0, \qquad \partial_- C_{+\alpha} = 0.
\label{e:11}
\end{equation}

In the case of other complex classical Lie group using conjugations we
always bring a $\bbZ_M$-gradation under consideration to the form
described by Figure \ref{f:1}. Here the mappings $c_+$ and $c_-$ take
the form described by Figures \ref{f:2} and \ref{f:3}, respectively,
where the mappings $C_{+\alpha}$ and $C_{-\alpha}$ satisfy the
relations (\ref{e:11}). Therefore, we call the structure of the
mappings $c_+$ and $c_-$ described by Figures \ref{f:2} and \ref{f:3}
canonical.

Return to the case of the complex general linear groups and
parameterize the mapping $\gamma$ as
\begin{equation}
\psset{xunit=2.2em, yunit=1.4em}
\gamma = \left( \begin{pspicture}[.4](.7,.8)(4.4,4.4)
\rput(1,4){$\Gamma_1$} \rput(2,3){$\Gamma_2$} \psddots(3,2)
\rput(4,1){$\Gamma_p$}
\end{pspicture} \right), \label{e:12}
\end{equation}
where for each $\alpha = 1, \ldots, p$ the mapping $\Gamma_\alpha$ is
a mapping from $\calM$ to the Lie group
$\mathrm{GL}_{n_\alpha}(\mathbb C)$. In this pa\-ra\-met\-rization the
Toda equation (\ref{e:5}) for the mapping $\gamma$ is equivalent to
the following system of equations for the mappings
$\Gamma_\alpha$:
\begin{align}
\partial_+ \left( \Gamma_1^{-1} \, \partial_- \Gamma_1^{} \right) &= -
\Gamma_1^{-1} C_{+1}^{} \, \Gamma_2^{} \, C_{-1}^{} + C_{-0}^{}
\Gamma_p^{-1} C_{+0}^{} \Gamma_1^{}, \notag \\*
\partial_+ \left( \Gamma_2^{-1} \, \partial_- \Gamma_2^{} \right) &= -
\Gamma_2^{-1} C_{+2}^{} \, \Gamma_3^{} \, C_{-2}^{} + C_{-1}^{}
\Gamma_1^{-1} C_{+1}^{} \Gamma_2^{}, \notag \\*
& \quad \vdots \label{e:13} \\*
\partial_+ \left( \Gamma_{p-1}^{-1} \, \partial_- \Gamma_{p-1}^{}
\right) &= - \Gamma_{p-1}^{-1} C_{+(p-1)}^{} \, \Gamma_p^{} \,
C_{-(p-1)}^{} + C_{-(p-2)}^{} \Gamma_{p-2}^{-1} C_{+(p-2)}^{}
\Gamma_{p-1}^{}, \notag \\*
\partial_+ \left( \Gamma_p^{-1} \, \partial_- \Gamma_p^{} \right) &= -
\Gamma_p^{-1} C_{+0}^{} \, \Gamma_1^{} \, C_{-0}^{} + C_{-(p-1)}^{}
\Gamma_{p-1}^{-1} C_{+(p-1)}^{} \Gamma_p^{}. \notag
\end{align}
If for some $\alpha$ one has $C_{+\alpha} = 0$ or $C_{-\alpha} = 0$,
then the system of equations (\ref{e:13}) is actually the system of
equations which arises when we consider a Toda equation associated
with the Lie group $\mathrm{GL}_n(\bbC)$, see, for example, the papers
\cite{RazSavZue99,NirRaz02}. Hence, to have equations which are really
associated with a loop group of $\rmGL_n(\bbC)$ one must assume that
all mappings $C_{+\alpha}$ and $C_{-\alpha}$ are nontrivial. This is
possible only when $k_\alpha = L$ and $M = p L$. Actually without any
loss of generality one can assume that $L = 1$.

It is often convenient to suppose that the mappings $\Gamma_\alpha$,
$C_{+\alpha}$ and $C_{-\alpha}$ are defined for all integer values of
the index $\alpha$ and satisfy the periodicity condition
\begin{gather*}
\Gamma_{\alpha+p} = \Gamma_\alpha, \\
C_{-(\alpha+p)} = C_{-\alpha}, \qquad C_{+(\alpha+p)} = C_{+\alpha}.
\end{gather*}
Now the system (\ref{e:13}) can be treated as the infinite periodic
system
\[
\partial_+ (\Gamma^{-1}_\alpha \partial_- \Gamma^{}_\alpha) = -
\Gamma^{-1}_\alpha C_{+\alpha} \Gamma^{}_{\alpha+1} C_{-\alpha} +
C_{-(\alpha-1)} \Gamma^{-1}_{\alpha-1} C_{+(\alpha-1)} \,
\Gamma^{}_\alpha.
\]
In particular, when $n = r p$ and $n_\alpha = r$, $C_{-\alpha} =
C_{+\alpha} = I_r$ one comes to the system
\[
\partial_+ (\Gamma^{-1}_\alpha \partial_- \Gamma^{}_\alpha) = -
\Gamma^{-1}_\alpha \Gamma^{}_{\alpha+1} + \Gamma^{-1}_{\alpha-1}
\Gamma^{}_\alpha.
\]

It is not difficult to understand that the Toda equations associated
with loop groups of $\rmSL_n(\bbC)$ in the case of $\bbZ_M$-gradations
of inner type have the same form (\ref{e:13}) as the Toda equations
associated with loop groups of $\rmGL_n(\bbC)$. Here the mappings
$\Gamma_\alpha$ should satisfy the condition $\prod_{\alpha=1}^p \det
\Gamma_\alpha = 1$. Actually, if there is a solution of a Toda
equation associated with a loop group of $\rmGL_n(\bbC)$ one can
easily obtain a solution of the corresponding Toda equation associated
with the corresponding loop group of $\rmSL_n(\bbC)$. Indeed, suppose
that the mappings $\Gamma_\alpha$, $ \alpha = 1, \ldots, p$, satisfy
the system (\ref{e:13}). Let us denote
\[
\Delta = \prod_{\alpha=1}^p \det \Gamma_\alpha.
\]
It is evident that
\[
\partial_+(\Delta^{-1} \partial_- \Delta) = \sum_{\alpha=1}^p
\partial_+[(\det \Gamma_\alpha)^{-1} \partial_- \det \Gamma_\alpha].
\]
The well known formula for the differentiation of a determinant gives
\[
\partial_+ [(\det \Gamma_\alpha)^{-1} \partial_- \det \Gamma_\alpha] =
\tr [ \partial_+( \Gamma_\alpha^{-1} \partial_- \Gamma_\alpha)].
\]
Taking into account the system (\ref{e:13}), we obtain
\[
\partial_+(\Delta^{-1} \partial_- \Delta) = \sum_{\alpha=1}^p \tr [
\partial_+( \Gamma_\alpha^{-1} \partial_- \Gamma_\alpha)] = 0.
\]
Now one can show that the mappings
\[
\widetilde \Gamma_\alpha = \Delta^{-1/n} \Gamma_\alpha, \qquad \alpha
= 1, \ldots, p,
\]
satisfy the condition $\prod_{\alpha=1}^p \det \widetilde
\Gamma_\alpha = 1$ and the system (\ref{e:13}). Note that every
solution of a Toda equation associated with a loop group of
$\rmSL_n(\bbC)$ can be obtained from a solution of the corresponding
Toda equation associated with the corresponding loop group
of~$\rmGL_n(\bbC)$.

The Toda equations arising when we use $\bbZ_M$-gradations of
$\gothgl_n(\bbC)$ of outer type, are considered in Section \ref{s:6}.

\section{Toda equations associated with loop groups of complex
orthogonal groups. Gradations of inner type} \label{s:4}

Let $a$ be an inner automorphism of the Lie group $\rmSO_n(\bbC)$
satisfying the relation $a^M = \id_{\rmSO_n(\bbC)}$. The corresponding
inner automorphism $A$ of the Lie algebra $\gothso_n(\bbC)$ satisfies
the relation $A^M = \id_{\gothso_n(\bbC)}$. In the simplest case $A =
\id_{\gothso_n(\bbC)}$ and $M$ is an arbitrary positive integer. In
this case we come to Eq.~(\ref{e:5a}), where $\Gamma$ is a
mapping from $\calM$ to $\rmSO_n(\bbC)$, $C_+$ and $C_-$ are mappings
from $\calM$ to $\gothso_n(\bbC)$ satisfying the
conditions~(\ref{e:6a}).

In a general case without any loss of generality we can assume that
the automorphism $A$ is a conjugation by an element $h$ which belongs
to the Lie group $\rmSO_n(\bbC) \cap \rmD_n(\bbC)$, see Appendix
\ref{a:c3}.

For any element $x \in \gothso_n(\bbC)$ one has $h^M \, x \, h^{-M} =
x$, therefore,
\[
h^M = \nu I
\]
for some complex number $\nu$. This equality implies that $({}^{J\!}
h)^M = \nu I$, therefore, $({}^{J\!} h)^M \, h^M = \nu^2 I$. From the
other hand, using the equality ${}^{J\!} h \, h = I$, we obtain
$({}^{J\!} h)^M \, h^M = ({}^{J\!} h \, h)^M = I$. Thus, we see that
$\nu^2 = 1$. In other words, either $\nu = 1$, or $\nu = -1$.

\subsection{\mathversion{bold}$\nu = 1$}

In this case $h$ is a diagonal matrix with the diagonal matrix
elements of the form $\epsilon_M^m$, where we assume that $m$ is a
nonnegative integer satisfying the conditions $0 < m \le M$. Let
$m_\alpha$, $\alpha = 1, \ldots, p$, be the different values of $m$
taken in the decreasing order, $M \ge m_1 > m_2 > \ldots > m_p > 0$,
and $n_\alpha$, $\alpha = 1, \ldots, p$, be their multiplicities. In
other words, for each $\alpha$ the element $h$ has $n_\alpha$ diagonal
elements equal to~$\epsilon_M^{m_\alpha}$.

As follows from the equality ${}^{J\!} h \, h = I$, if the diagonal
element of $h$ at position $i$ is $\epsilon_M^m$, then the diagonal
element of $h$ at position $n-i+1$ is equal to $\epsilon_M^{M-m}$.
Note that if $n$ is odd, then the equality $\det h = 1$ implies that
the central diagonal matrix element of $h$ is equal to $1$.

Below, to bring the element $h$ into the desirable form, we use two
special types of automorphisms of the Lie group $\mathrm{SO}_n(\bbC)$.
The automorphisms of the first type are conjugations by the matrices
of the following block matrix form
\begin{equation}
\left(\begin{array}{ccccc}
I_{i-1} & & & & \\
& 0 & 0 & 1 & \\
& 0 & I_{n - 2i} & 0 & \\
& 1 & 0 & 0 & \\
& & & & I_{i-1}
\end{array}
\right), \label{e:14}
\end{equation}
where $i \le n/2$. The corresponding automorphism permutes the
diagonal elements of a diagonal matrix at positions $i$ and $n-i+1$.
The automorphisms of the second type are conjugations by the matrices
of the form
\begin{equation}
\left(\begin{array}{ccccccc}
I_{i-1} & & & & & & \\
& 0 & 1 & & & & \\
& 1 & 0 & & & & \\
& & & I_{n-2i-2} & & & \\
& & & & 0 & 1 & \\
& & & & 1 & 0 & \\
& & & & & & I_{i-1}
\end{array}
\right), \label{e:15}
\end{equation}
where $i \le n/2-1$. The corresponding automorphism of $\rmSO_n(\bbC)$
permutes the diagonal elements of a diagonal matrix at positions $i$
and $i+1$ and the diagonal elements at positions $n-i+1$ and $n-i$.

\subsubsection{\mathversion{bold}$m_1 = M$} \label{s:4.1.1}

Assume now that $m_1 = M$. It is not difficult to see that $n-n_1$ is
an even integer. Using automorphisms generated by elements of the type
(\ref{e:14}) and of the type (\ref{e:15}), we bring the element $h$ to
the form
\[
h = \left( \begin{array}{ccc}
h' & & \\
& I_{n_1} & \\
& & ({}^{J\!} h')^{-1}
\end{array}
\right),
\]
where $h'$ is a square diagonal matrix of order $(n-n_1)/2$ which
satisfies the relation $h^{\prime M} = I_{(n-n_1)/2}$ and the
condition that the diagonal matrix elements of $h'$ have the form
$\epsilon_M^m$, where $M > m \ge M/2$, and $m$ does not increase when
we go along the diagonal from the upper left corner to the bottom
right one.

The conjugation by the matrix
\begin{equation}
\left(\begin{array}{ccc}
0 & I_{n_1} & \\
I_{(n-n_1)/2} & 0 & \\
& & I_{(n-n_1)/2} \label{e:16}
\end{array}
\right)
\end{equation}
maps $\gothso_n(\bbC)$ isomorphically to the Lie algebra
$\gothgl_n^B(\bbC)$ with
\begin{equation}
B = \left( \begin{array}{cc}
J_{n_1} & 0 \\
0 & J_{n-n_1}
\end{array} \right). \label{e:17}
\end{equation}
The same conjugation maps the Lie group $\rmO_n(\bbC)$ isomorphically
to the Lie group $\rmGL_n^B(\bbC)$. Here the Lie group $\rmSO_n(\bbC)$
is mapped isomorphically to the Lie subgroup of $\rmGL_n^B(\bbC)$
formed by the elements $g$ of $\rmGL_n^B(\bbC)$ with $\det g = 1$.

Now the element $h$ takes the form
\[
h = \left( \begin{array}{cc}
I_{n_1} & \\
 & h''
\end{array}\right),
\]
where
\[
h'' = \left( \begin{array}{cc}
h' & \\
& ({}^{J\!} h')^{-1}
\end{array} \right).
\]
Note that ${}^{J\!} h'' \, h'' = I_{n-n_1}$. Using automorphisms of
$\rmGL_n^B(\bbC)$ similar to automorphisms of $\rmSO_n(\bbC)$
generated by the elements of the form (\ref{e:14}) or (\ref{e:15}), we
reduce $h$ to the form (\ref{e:7}), where the numbers $n_\alpha$
satisfy the relations
\begin{equation}
n_\alpha = n_{p-\alpha+2}, \qquad \alpha = 2, \ldots, p. \label{e:18}
\end{equation}
It is clear that $\sum_{\alpha=1}^p n_\alpha = n$. The numbers
$\mu_\alpha$ satisfy the equalities $\mu_\alpha^M = 1$, $\alpha = 1,
\ldots, p$, and
\begin{equation}
\mu_\alpha = \mu_{p-\alpha+2}^{-1}, \qquad \alpha = 2, \ldots, p.
\label{e:19}
\end{equation}
The integers $m_\alpha$ form a decreasing sequence and satisfy the
relations
\begin{equation}
m_\alpha = M - m_{p-\alpha+2}, \qquad \alpha = 2, \ldots, p.
\label{e:20}
\end{equation}
The positive integers $k_\alpha$, $\alpha = 1, \ldots, p-1$, defined
as $k_\alpha = m_\alpha - m_{\alpha+1}$, satisfy the relations
\begin{equation}
k_\alpha = k_{p-\alpha+1}, \qquad \alpha = 2, \ldots, p-1.
\label{e:21}
\end{equation}
One, certainly, has
\[
m_\alpha = \sum_{\beta = \alpha}^{p-1} k_\beta + m_p.
\]
Using (\ref{e:20}), we see that $m_p = M - m_2 = m_1 - m_2 = k_1$.
Hence, we come to the equality
\begin{equation}
m_\alpha = \sum_{\beta = \alpha}^{p-1} k_\beta + k_1. \label{e:22}
\end{equation}
In particular, remembering that $m_1 = M$, we obtain
\begin{equation}
\sum_{\beta=1}^{p-1} k_\beta + k_1 = M. \label{e:23}
\end{equation}

Assume now that there are given $p$ positive integers $n_\alpha$ such
that $\sum_{\alpha=1}^p n_\alpha = n$, the integer $n-n_1$ is even,
and the relations (\ref{e:18}) are satisfied. Assume also that there
are given $p-1$ positive integers $k_\alpha$ satisfying the relations
(\ref{e:21}) and the equality (\ref{e:23}). Using the relation
(\ref{e:22}), we define $p$ positive integers $m_\alpha$. Here the
relation (\ref{e:23}) guarantees that $m_1 = M$. It follows from
(\ref{e:21}) and (\ref{e:23}) that the integers $m_\alpha$ satisfy the
relations (\ref{e:20}). Hence, the numbers $\mu_\alpha = \epsilon_M^{m_\alpha}$
satisfy the relations (\ref{e:19}). The matrix $h$ defined by
(\ref{e:7}) belongs to the Lie group $\rmGL_n^B(\bbC)$ with $B$
defined by (\ref{e:17}). The element $h$ satisfies the equality $h^M =
I$.

Thus, a $\bbZ_M$-gradation of $\gothso_n(\bbC)$ for which $\nu = 1$
and $m_1 = M$ is specified by a choice of $p$ positive integers
$n_\alpha$ such that $\sum_{\alpha=1}^p n_\alpha = n$, the integer
$n-n_1$ is even, and the relations (\ref{e:18}) are satisfied, and by
a choice of $p-1$ positive integers $k_\alpha$ satisfying the
relations (\ref{e:21}) and the equality (\ref{e:23}).

The structure of the $\bbZ_M$-gradation generated by the automorphism
$A$ under consideration is again described by Figure~\ref{f:1}. Here
an element $x$ of the Lie algebra $\gothgl_n^B(\bbC)$ is treated as a
matrix having the block matrix structure (\ref{e:10}). Now the blocks
$x_{\alpha \beta}$ are not arbitrary. They satisfy the restrictions
which follow from the equality ${}^{B\!} x = -x$. The Lie subalgebra
$\gothg_{[0]_M}$ is formed by block diagonal matrices. It is clear
that the Lie group $G_0$ is also formed by block diagonal matrices. It
is the desire to have such a simple structure of $\gothg_{[0]_M}$ and
$G_0$ that is the reason to use the Lie algebra $\gothgl_n^B(\bbC)$
and a Lie subgroup of the Lie group $\rmGL_n^B(\bbC)$ with $B$ defined
by the relation (\ref{e:17}) instead of the Lie algebra
$\gothso_n(\bbC)$ and the Lie group $\rmSO_n(\bbC)$.

Below, using conjugations, we always bring $\gothg_{[0]_M}$ and $G_0$
to block diagonal form. It allows one to use for the mapping $\gamma$,
entering the Toda equation (\ref{e:5}), the parametrization
(\ref{e:12}). Here, in general, the mappings $\Gamma_\alpha$, $\alpha
= 1, \ldots, p$, are not independent. We denote the number of
independent mappings $\Gamma_\alpha$ uniquely determining the mapping
$\gamma$ by $s$. In the cases considered below it is always possible
to choose the first $s$ mappings $\Gamma_\alpha$ as a complete set of
independent mappings.

Describe now the Toda equations arising in the case where $\nu = 1$
and $m_1 = M$. Choose some $\bbZ_M$-gradation of the Lie algebra
$\gothso_n(\bbC)$ of the type under consideration. Then map
$\gothso_n(\bbC)$ isomorphically onto the Lie algebra $\gothg =
\gothgl_n^B(\bbC)$ with $B$ given by~(\ref{e:17}), and consider the
corresponding conjugated $\bbZ_M$-gradation of $\gothgl_n^B(\bbC)$.
Let $L$ be a positive integer such that the grading subspace
$\gothg_{[k]_M}$ is trivial if $0 < |k| < L$. One can get convinced
that if $x \in \gothg_{+[L]_M}$, then only the blocks $x_{\alpha,
\alpha+1}$, $\alpha = 1, \ldots, p-1$, and $x_{p1}$ in the block
matrix representation (\ref{e:10}) can be different from zero.
Similarly, if $x \in \gothg_{-[L]_M}$, then only the blocks $x_{\alpha
+ 1, \alpha}$, $\alpha = 1, \ldots, p-1$, and $x_{1p}$ can be
different from zero. It is convenient to consider the cases of an even
and odd $p$ separately.

First let $p$ be an even integer equal to $2s-2$, $s \ge 2$. It
follows from (\ref{e:23}) and (\ref{e:21}) that this is possible only
if $M$ is even. Note that the integer $n_s$ is also even in this case.
Having in mind the restrictions which follow from the equality
${}^{B\!} x = -x$, we see that the mapping $c_+$ may be parameterized
as it is given in Figure \ref{f:2}, where in our case $C_{+0} = -
{}^{J\!} C_{+1}$, and $C_{+\alpha} = - {}^{J\!} C_{+(p-\alpha+1)}$,
$\alpha = 2, \ldots, p-1$. Here we again assume that if some blocks
among $x_{\alpha, \alpha+1}$, $\alpha = 1, \ldots, p-1$, and $x_{p1}$
in the general block matrix representation (\ref{e:10}) have the
grading index different from $[L]_M$, then the corresponding blocks in
the block representation of $c_+$ are zero matrices. A similar
convention is used for the block representation of the mapping $c_-$,
which we choose in the form given in Figure \ref{f:3}, where in our
case $C_{-0} = - {}^{J\!} C_{-1}$, and $C_{-\alpha} = - {}^{J\!}
C_{-(p-\alpha+1)}$, $\alpha = 2, \ldots, p-1$.

Parameterize the mapping $\gamma$ as is described by the equality
(\ref{e:12}). Here $\Gamma_1 = ({}^{J\!} \Gamma_1)^{-1}$ and
$\Gamma_\alpha = ({}^{J\!} \Gamma_{p-\alpha+2})^{-1}$ for each $\alpha
= 2, \ldots, p$. It follows from the last relation that $\Gamma_s =
({}^{J\!} \Gamma_s)^{-1}$. Therefore, among the mappings
$\Gamma_\alpha$ there are only $s$ independent ones, and the Lie group
$G_0$ is isomorphic to the Lie group $\rmSO_{n_1}(\bbC) \times
\rmGL_{n_2}(\bbC) \times \cdots \times \rmGL_{n_{s-1}}(\bbC) \times
\rmSO_{n_s}(\bbC)$. Recall that in the case under consideration $n_s$
is even. One can get convinced that the Toda equation (\ref{e:5}) is
equivalent to the following system of equations
\begin{align}
\partial_+ \left( \Gamma_1^{-1} \, \partial_- \Gamma_1^{} \right) &= -
\Gamma_1^{-1} C_{+1}^{} \, \Gamma_2^{} \, C_{-1}^{} + {}^{J\!}
(\Gamma_1^{-1}
C_{+1}^{} \, \Gamma_2^{} \, C_{-1}^{}), \notag \\*
\partial_+ \left( \Gamma_2^{-1} \, \partial_- \Gamma_2^{} \right) &= -
\Gamma_2^{-1} C_{+2}^{} \, \Gamma_3^{} \, C_{-2}^{} + C_{-1}^{}
\Gamma_1^{-1} C_{+1}^{} \Gamma_2^{}, \notag \\*
& \quad \vdots \label{e:24} \\*
\partial_+ \left( \Gamma_{s-1}^{-1} \, \partial_- \Gamma_{s-1}^{}
\right)
&= - \Gamma_{s-1}^{-1} C_{+(s-1)}^{} \, \Gamma_s^{} \, C_{-(s-1)}^{} +
C_{-(s-2)}^{} \Gamma_{s-2}^{-1} C_{+(s-2)}^{} \Gamma_{s-1}^{}, \notag
\\*
\partial_+ \left( \Gamma_s^{-1} \, \partial_- \Gamma_s^{} \right) &= -
{}^{J\!} (C_{-(s-1)}^{} \Gamma_{s-1}^{-1} C_{+(s-1)}^{} \Gamma_s^{}) +
C_{-(s-1)}^{} \Gamma_{s-1}^{-1} C_{+(s-1)}^{} \Gamma_s^{}. \notag
\end{align}

Let now $p$ be an odd integer equal to $2s-1$, $s \ge 2$. Parameterize
the mapping $c_+$ as is given in Figure \ref{f:2}, where $C_{+0} = -
{}^{J\!} C_{+1}$, and $C_{+\alpha} = - {}^{J\!} C_{+(p-\alpha+1)}$,
$\alpha = 2, \ldots, p-1$. For the mapping $c_-$ we use the
representation given in Figure \ref{f:3}, where $C_{-0} = - {}^{J\!}
C_{-1}$, and $C_{-\alpha} = - {}^{J\!} C_{-(p-\alpha+1)}$, $\alpha =
2, \ldots, p-1$.

Parameterize again the mapping $\gamma$ as is described by the
equality (\ref{e:12}).
Here $\Gamma_1 = ({}^{J\!} \Gamma_1)^{-1}$ and $\Gamma_\alpha =
({}^{J\!} \Gamma_{p-\alpha+2})^{-1}$ for each $\alpha = 2, \ldots, p$.
It is clear that there are only $s$ independent mappings
$\Gamma_\alpha$ and the Lie group $G_0$ is isomorphic to the Lie group
$\rmSO_{n_1}(\bbC) \times \rmGL_{n_2}(\bbC) \times \cdots \times
\rmGL_{n_s}(\bbC)$. Now we see that in the case under consideration
the following system of equations
\begin{align}
\partial_+ \left( \Gamma_1^{-1} \, \partial_- \Gamma_1^{} \right) &= -
\Gamma_1^{-1} C_{+1}^{} \, \Gamma_2^{} \, C_{-1}^{} + {}^{J\!}
(\Gamma_1^{-1}
C_{+1}^{} \, \Gamma_2^{} \, C_{-1}^{}), \notag \\*
\partial_+ \left( \Gamma_2^{-1} \, \partial_- \Gamma_2^{} \right) &= -
\Gamma_2^{-1} C_{+2}^{} \, \Gamma_3^{} \, C_{-2}^{} + C_{-1}^{}
\Gamma_1^{-1} C_{+1}^{} \Gamma_2^{}, \notag \\*
& \quad \vdots \label{e:25} \\*
\partial_+ \left( \Gamma_{s-1}^{-1} \, \partial_- \Gamma_{s-1}^{}
\right)
&= - \Gamma_{s-1}^{-1} C_{+(s-1)}^{} \, \Gamma_s^{} \, C_{-(s-1)}^{} +
C_{-(s-2)}^{} \Gamma_{s-2}^{-1} C_{+(s-2)}^{} \Gamma_{s-1}^{}, \notag
\\*
\partial_+ \left( \Gamma_s^{-1} \, \partial_- \Gamma_s^{} \right) &= -
\Gamma_s^{-1} C_{+s}^{} {}^{J\!} (\Gamma_s^{-1}) C_{-s}^{} +
C_{-(s-1)}^{} \Gamma_{s-1}^{-1} C_{+(s-1)}^{} \Gamma_s^{} \notag
\end{align}
is equivalent to the Toda equation~(\ref{e:5}). Note that $C_{+s} = -
{}^{J\!} C_{+s}$ and $C_{-s} = - {}^{J\!} C_{-s}$.

As well as for the case of Toda equations associated with loop groups
of complex general linear groups, to have equations which cannot be
reduced to Toda equations associated with the corresponding finite
dimensional groups, we should require all mappings $C_{+\alpha}$ and
$C_{-\alpha}$ entering the systems (\ref{e:25}) and (\ref{e:24}) to be
nontrivial. This is again possible only if we choose $k_\alpha = L$
for each $\alpha$ and $M = p L$.

\subsubsection{\mathversion{bold}$m_1 < M$} \label{s:4.1.2}

As we noted above, if $n$ is odd, then the equality $\det h = 1$
implies that the central diagonal matrix element of $h$ is $1$. When
$m_1 < M$ it is impossible. Therefore, in the case under consideration
$n$ is even. Using automorphisms of $\gothso_n(\bbC)$, generated by
elements of the form (\ref{e:14}) and of the form (\ref{e:15}), we
bring the element $h$ to the form (\ref{e:7}), where $n_\alpha$,
$\alpha = 1, \ldots, p$, are positive integers, such that
$\sum_{\alpha=1}^p n_\alpha = n$, and $\mu_\alpha$ are complex
numbers, such that $\mu_\alpha^M = 1$ for each $\alpha = 1, \ldots,
p$.

It follows from the equality ${}^{J\!} h \, h = I_n$ that the integers
$n_\alpha$
satisfy the relations
\begin{equation}
n_\alpha = n_{p-\alpha+1}, \qquad \alpha = 1, \ldots, p, \label{e:26}
\end{equation}
and the numbers $\mu_\alpha$ satisfy the relations
\begin{equation}
\mu_\alpha = \mu_{p-\alpha+1}^{-1}, \qquad \alpha = 1, \ldots, p.
\label{e:27}
\end{equation}
For the corresponding integers $m_\alpha$ one obtains
\begin{equation}
m_\alpha = M - m_{p-\alpha+1}, \qquad \alpha = 1, \ldots, p,
\label{e:28}
\end{equation}
that implies that the integers $k_\alpha$, $\alpha = 1, \ldots, p-1$,
defined as $k_\alpha = m_\alpha - m_{\alpha+1}$, satisfy the
equalities
\begin{equation}
k_\alpha = k_{p-\alpha}, \qquad \alpha = 1, \ldots, p-1. \label{e:29}
\end{equation}
Using the relations
\begin{equation}
m_\alpha = \sum_{\beta = \alpha}^{p-1} k_\beta + m_p, \qquad \alpha =
1,
\ldots, p-1, \label{e:30}
\end{equation}
and (\ref{e:28}), one obtains
\[
m_p = M - m_1 = M - \sum_{\beta=1}^{p-1} k_\beta - m_p.
\]
Hence, we come to the equality
\[
2 m_p = M - \sum_{\beta=1}^{p-1} k_\beta,
\]
which implies that $M - \sum_{\beta=1}^{p-1} k_\beta$ is an even
positive
integer.

Let now $n$ be an even positive integer, and $M$ be a positive
integer. Assume that there are given $p$ positive integers $n_\alpha$
satisfying the equality $\sum_{\alpha=1}^p n_\alpha = n$ and the
relations~(\ref{e:26}). Assume also that there are given $p-1$
positive integers $k_\alpha$ such that $M - \sum_{\beta=1}^{p-1}
k_\beta$ is an even positive integer, and which satisfy the relations
(\ref{e:29}). Putting
\[
m_p = \frac{1}{2} \left( M - \sum_{\beta=1}^{p-1} k_\beta \right),
\]
and using the relation (\ref{e:30}), we obtain $p$ integers $m_\alpha$
which satisfy the relations (\ref{e:28}). In this case the numbers
$\mu_\alpha = \epsilon_M^{m_\alpha}$ satisfy the relations
(\ref{e:27}) and the matrix $h$, defined by the equality (\ref{e:7}),
belongs to $\rmSO_n(\bbC) \cap \rmD_n(\bbC)$. Here we have $h^M = I$
and $m_1 < M$.

Thus, for an even $n$ a $\bbZ_M$-gradation of inner type of the Lie
algebra $\gothso_n(\bbC)$ for which $\nu = 1$ and $m_1 < M$ is
specified by a choice of positive integers $n_\alpha$ satisfying the
equality $\sum_{\alpha=1}^p n_\alpha = n$ and the relations
(\ref{e:26}), and positive integers $k_\alpha$ such that $M -
\sum_{\beta=1}^{p-1} k_\beta$ is an even positive integer and which
satisfy the relations (\ref{e:29}).

Proceed now to the corresponding Toda equations. Choose some
$\bbZ_M$-gradation of the Lie algebra $\gothso_n(\bbC)$ of the type
under consideration. Let $L$ be a positive integer such that the
grading subspace $\gothg_{[k]_M}$ is trivial if $0 < |k| < L$. As
above, if $x \in \gothg_{+[L]_M}$, then only the blocks $x_{\alpha,
\alpha+1}$, $\alpha = 1, \ldots, p-1$, and $x_{p1}$ in the block
matrix representation (\ref{e:10}) can be different from zero, and if
$x \in \gothg_{-[L]_M}$, then only the blocks $x_{\alpha + 1,
\alpha}$, $\alpha = 1, \ldots, p-1$, and $x_{1p}$ can be different
from zero.

Let $p$ be an odd integer equal to $2s-1$, $s \ge 2$. Having in mind
the restrictions which follow from the equality ${}^{J\!} x = - x$,
parameterize the mapping $c_+$ as is given in Figure \ref{f:2}, where
$C_{+\alpha} = - {}^{J\!} C_{+(p-\alpha)}$, $\alpha = 1, \ldots, p-1$,
and $C_{+0} = - {}^{J\!} C_{+0}$. For the mapping $c_-$ we use the
parametrization given in Figure \ref{f:3}, where $C_{-\alpha} = -
{}^{J\!} C_{-(p-\alpha)}$, $\alpha = 1, \ldots, p-1$, and $C_{-0} = -
{}^{J\!} C_{-0}$.

Parameterize the mapping $\gamma$ in accordance with the relation
(\ref{e:12}). Here for each $\alpha$ we have $\Gamma_{p-\alpha+1} =
({}^{J\!} \Gamma_{\alpha})^{-1}$. It is clear that $\Gamma_s =
({}^{J\!} \Gamma_s)^{-1}$, and that the Lie group $G_0$ is isomorphic
to the Lie group $\rmGL_{n_1}(\bbC) \times \cdots \times
\rmGL_{n_{s-1}}(\bbC) \times \rmSO_{n_s}(\bbC)$. We come to the
following system of equations
\begin{align}
\partial_+ \left( \Gamma_1^{-1} \, \partial_- \Gamma_1^{} \right) &= -
\Gamma_1^{-1} C_{+1}^{} \, \Gamma_2^{} \, C_{-1}^{} + C_{-0}^{} \,
{}^{J\!} \Gamma_1 \, C_{+0}^{} \Gamma_1^{}, \notag \\*
\partial_+ \left( \Gamma_2^{-1} \, \partial_- \Gamma_2^{} \right) &= -
\Gamma_2^{-1} C_{+2}^{} \, \Gamma_3^{} \, C_{-2}^{} + C_{-1}^{}
\Gamma_1^{-1} C_{+1}^{} \Gamma_2^{}, \notag \\*
& \quad \vdots \label{e:31} \\*
\partial_+ \left( \Gamma_{s-1}^{-1} \, \partial_- \Gamma_{s-1}^{}
\right)
&= - \Gamma_{s-1}^{-1} C_{+(s-1)}^{} \, \Gamma_s^{} \, C_{-(s-1)}^{} +
C_{-(s-2)}^{} \Gamma_{s-2}^{-1} C_{+(s-2)}^{} \Gamma_{s-1}^{}, \notag
\\*
\partial_+ \left( \Gamma_s^{-1} \, \partial_- \Gamma_s^{} \right) &= -
{}^{J\!} (C_{-(s-1)}^{} \Gamma_{s-1}^{-1} C_{+(s-1)}^{} \Gamma_s^{}) +
C_{-(s-1)}^{} \Gamma_{s-1}^{-1} C_{+(s-1)}^{} \Gamma_s^{} \notag
\end{align}
which is equivalent to the Toda equation (\ref{e:5}). Actually the
substitution $\Gamma_\alpha \to {}^{J\!}(\Gamma_{s-\alpha+1}^{-1})$
and $C_{\pm \alpha} \to - {}^{J\!} C_{\pm (s-\alpha)}$ transforms the
system (\ref{e:31}) into the system (\ref{e:25}). Hence in the case
under consideration we do not obtain new Toda equations.

Let now $p$ be an even integer equal to $2s$, $s \ge 1$. Parameterize
the mapping $c_+$ as is given in Figure \ref{f:2}, where $C_{+\alpha}
= - {}^{J\!} C_{+(p-\alpha)}$, $\alpha = 1, \ldots, p-1$, and $C_{+0}
= - {}^{J\!} C_{+0}$. For the mapping $c_-$ we use the parametrization
given in Figure \ref{f:3}, where $C_{-\alpha} = - {}^{J\!}
C_{-(p-\alpha)}$, $\alpha = 1, \ldots, p-1$, and $C_{-0} = - {}^{J\!}
C_{-0}$. An appropriate parametrization of the mapping $\gamma$ is
given by the equality (\ref{e:12}), where for each $\alpha = 1,
\ldots, p$ we have $\Gamma_{p-\alpha+1} = ({}^{J\!}
\Gamma_{\alpha})^{-1}$. The Lie group $G_0$ is isomorphic to the Lie
group $\rmGL_{n_1}(\bbC) \times \cdots \times \rmGL_{n_s}(\bbC)$, and
the Toda equation (\ref{e:5}) is equivalent in the case under
consideration to the system
\begin{align}
\partial_+ \left( \Gamma_1^{-1} \, \partial_- \Gamma_1^{} \right) &= -
\Gamma_1^{-1} C_{+1}^{} \, \Gamma_2^{} \, C_{-1}^{} + C_{-0}^{} \,
{}^{J\!} \Gamma_1 \, C_{+0}^{} \Gamma_1^{}, \notag \\*
\partial_+ \left( \Gamma_2^{-1} \, \partial_- \Gamma_2^{} \right) &= -
\Gamma_2^{-1} C_{+2}^{} \, \Gamma_3^{} \, C_{-2}^{} + C_{-1}^{}
\Gamma_1^{-1} C_{+1}^{} \Gamma_2^{}, \notag \\*
& \quad \vdots \label{e:32} \\*
\partial_+ \left( \Gamma_{s-1}^{-1} \, \partial_- \Gamma_{s-1}^{}
\right)
&= - \Gamma_{s-1}^{-1} C_{+(s-1)}^{} \, \Gamma_s^{} \, C_{-(s-1)}^{} +
C_{-(s-2)}^{} \Gamma_{s-2}^{-1} C_{+(s-2)}^{} \Gamma_{s-1}^{}, \notag
\\*
\partial_+ \left( \Gamma_s^{-1} \, \partial_- \Gamma_s^{} \right) &= -
\Gamma_s^{-1} C_{+s}^{} {}^{J\!} (\Gamma_s^{-1}) C_{-s} +
C_{-(s-1)}^{}
\Gamma_{s-1}^{-1} C_{+(s-1)}^{} \Gamma_s^{}. \notag
\end{align}

Only in the case $k_\alpha = L$ and $M = p L$ the systems (\ref{e:31})
and (\ref{e:32}) cannot be reduced to Toda equations associated with
the finite dimensional Lie groups $\rmSO_n(\bbC)$. Here the condition
that $M - \sum_{\beta=1}^{p-1} k_\beta$ is an even positive integer
implies that $L$ should be even.

\subsection{\mathversion{bold}$\nu = -1$} \label{s:4.2}

Since in the case under consideration $h^M = -I$ and $\det h = 1$, the
integer $n$ must be even. The matrix $h$ is a diagonal matrix with the
diagonal matrix elements of the form $\epsilon_M^m/\epsilon_{2M}$,
where we assume that $0 < m \le M$. As before, we denote by
$m_\alpha$, $\alpha = 1, \ldots, p$, the different values of $m$ taken
in decreasing order, and by $n_\alpha$, $\alpha = 1, \ldots, p$, their
multiplicities. Using automorphisms of $\gothso_n(\bbC)$, generated by
elements of the type (\ref{e:14}) and of the type (\ref{e:15}), we
bring the element $h$ to the form (\ref{e:7}).

It follows from the equality ${}^{J\!} h \, h = I$ that the integers
$n_\alpha$ satisfy the relations (\ref{e:26}), and the numbers
$\mu_\alpha$ satisfy the relations (\ref{e:27}). For the corresponding
integers $m_\alpha$ instead of (\ref{e:28}) one obtains the equalities
\begin{equation}
m_\alpha = M - m_{p-\alpha+1} + 1, \qquad \alpha = 1, \ldots, p.
\label{e:33}
\end{equation}
One can show that now we have the equality
\[
2 m_p = M - \sum_{\beta=1}^{p-1} k_\beta + 1,
\]
where the integers $k_\alpha$ are defined as $k_\alpha = m_\alpha -
m_{\alpha+1}$ as before. This equality implies that $M -
\sum_{\beta=1}^{p-1} k_\beta + 1$ is an even positive integer.

Let now $n$ be an even positive integer, and $M$ be a positive
integer. Assume that there are given $p$ positive integers $n_\alpha$
satisfying the equality $\sum_{\alpha=1}^p n_\alpha = n$ and the
relations (\ref{e:26}). Assume also that there are given $p-1$
positive integers $k_\alpha$ such that $M - \sum_{\beta=1}^{p-1}
k_\beta + 1$ is an even positive integer, and which satisfy the
relations (\ref{e:29}). Using the relation (\ref{e:30}) and putting
\[
m_p = \frac{1}{2} \left( M - \sum_{\beta=1}^{p-1} k_\beta + 1 \right),
\]
we obtain $p$ integers $m_\alpha$ which satisfy the relations
(\ref{e:33}). In this case the numbers $\mu_\alpha =
\epsilon_M^{m_\alpha}/\epsilon_{2M}$ satisfy the relations
(\ref{e:27}), and the matrix $h$ defined by (\ref{e:7}) satisfies the
equality ${}^{J\!} h \, h = I$. Moreover, $h$ satisfies the equalities
$\det h = 1$ and $h^M = - I$.

Thus, for an even $n$ a $\bbZ_M$-gradation of the Lie algebra
$\gothso_n(\bbC)$ of inner type for which $\nu = -1$ is specified by a
choice of positive integers $n_\alpha$ satisfying the equality
$\sum_{\alpha=1}^p n_\alpha = n$ and the relations (\ref{e:26}), and
positive integers $k_\alpha$ such that $M - \sum_{\beta=1}^{p-1}
k_\beta + 1$ is an even positive integer, and which satisfy the
relations (\ref{e:29}).

It is not difficult to see that using $\bbZ_M$-gradations of the type
under consideration we again come to the systems (\ref{e:31}) and
(\ref{e:32}). The only difference is that to have Toda equations which
cannot be reduced to Toda equations associated with the finite
dimensional Lie groups $\rmSO_n(\bbC)$ one uses $\bbZ_M$-gradations
for which the integer $L$ is odd.

\section{Toda equations associated with loop groups of complex
symplectic groups}

There are only inner automorphisms of the Lie group $\rmSp_n(\bbC)$
and the Lie algebra $\gothsp_n(\bbC)$. Let $a$ be an automorphism of
$\rmSp_n(\bbC)$ satisfying the relation $a^M = \id_{\rmSp_n(\bbC)}$.
The corresponding automorphism $A$ of the Lie algebra
$\gothsp_n(\bbC)$ satisfies the relation $A^M =
\id_{\gothsp_n(\bbC)}$. In the case when $A = \id_{\gothsp_n(\bbC)}$
and $M$ is an arbitrary positive integer we come to Eq.~(\ref{e:5a}),
where $\Gamma$ is a mapping from $\calM$ to
$\rmSp_n(\bbC)$, $C_+$ and $C_-$ are mappings from $\calM$ to
$\gothsp_n(\bbC)$ satisfying the conditions (\ref{e:6a}).

In a general case without any loss of generality we can assume that
the automorphism $A$ is given by the relation
\[
A(x) = h \, x \, h^{-1},
\]
where $h \in \rmSp_n(\bbC) \cap \rmD_n(\bbC)$, see Appendix
\ref{a:c4}. By definition we have ${}^{\!K} h = h^{-1}$. It is easy to
see that since $h \in \rmSp_n(\bbC) \cap \rmD_n(\bbC)$, one has
${}^{\!K} h = {}^{\!J} h$, therefore, $h \in \rmSO_n(\bbC) \cap
\rmD_n(\bbC)$. Furthermore, one has $h^M = \nu I$, where $\nu$ is
either $1$, or $-1$. Thus, the matrix $h$ satisfies the same relations
as for the case of complex orthogonal groups. Actually, to describe
Toda equations associated with loop groups of complex symplectic
groups, one can almost literally follow the lines of Section
\ref{s:4}.

\subsection{\mathversion{bold}$\nu = 1$}

Assume that $\nu = 1$. Here, as for the case of complex orthogonal
groups, $h$ is a diagonal matrix with the diagonal matrix elements of
the form $\epsilon_M^m$, where we assume that $m$ is a nonnegative
integer satisfying the conditions $0 < m \le M$. Let $m_\alpha$,
$\alpha = 1, \ldots, p$, be the different values of $m$ taken in the
decreasing order, $M \ge m_1 > m_2 > \ldots > m_p > 0$, and
$n_\alpha$, $\alpha = 1, \ldots, p$, be their multiplicities.

\subsubsection{\mathversion{bold}$m_1 = M$}

Suppose that $m_1 = M$. To bring the element $h$ to a form convenient
for our purposes we permute its diagonal matrix elements by two types
of automorphisms of $\gothsp_n(\bbC)$. The automorphisms of the first
type are conjugations by the matrices of the block matrix form
\[
\left(\begin{array}{ccccc}
I_{i-1} & & & & \\
& 0 & 0 & \rmi & \\
& 0 & I_{n - 2i} & 0 & \\
& \rmi & 0 & 0 & \\
& & & & I_{i-1}
\end{array}
\right),
\]
where $i \le n/2$. The corresponding automorphism permutes the
diagonal elements of a diagonal matrix at positions $i$ and $n-i+1$.
The automorphisms of the second type are conjugations by the matrices
of the form described by (\ref{e:15}), where $i \le n/2-1$. The
corresponding automorphism of $\gothsp_n(\bbC)$ permutes the diagonal
elements of a diagonal matrix at positions $i$ and $i+1$ and the
diagonal elements at positions $n-i+1$ and $n-i$. Using automorphisms
of these two types and the conjugation by the matrix of the form
(\ref{e:16}), we bring the element $h$ to the same form as in the case
of complex orthogonal groups, see Section \ref{s:4.1.1}. Here the
conjugation by the matrix of the form (\ref{e:16}) maps the Lie
algebra $\gothsp_n(\bbC)$ isomorphically to the Lie algebra
$\gothgl_n^B(\bbC)$ with
\[
B = \left( \begin{array}{cc}
K_{n_1} & 0 \\
0 & K_{n-n_1}
\end{array} \right).
\]
The same conjugation maps the Lie group $\rmSp_n(\bbC)$ isomorphically
to the Lie group $\rmGL_n^B(\bbC)$.

Repeating the discussion given in Section \ref{s:4.1.1}, we conclude
that in the case under consideration a $\bbZ_M$-gradation of the Lie
algebra $\gothsp_n(\bbC)$ is specified by a choice of $p$ positive
integers $n_\alpha$ such that $\sum_{\alpha=1}^p n_\alpha = n$,
satisfying the relations (\ref{e:18}), and a choice of $p-1$ positive
integers $k_\alpha$, satisfying the relations (\ref{e:21}) and the
equality (\ref{e:23}). The structure of the $\bbZ_M$-gradation is
described by Figure~\ref{f:1}.

Choose a $\bbZ_M$-gradation of the type under consideration and
describe the corresponding Toda equation. Consider first the case of
an even $p = 2s - 2$, $s \ge 2$. Using the restrictions which follow
from the equality ${}^{B\!} x = - x$, parameterize the mapping $c_+$
as is given in Figure \ref{f:2}, where $C_{+0} = {}^{JK\!} C_{+1}$,
$C_{+s} = -{}^{KJ\!} C_{+(s-1)}$, and $C_{+\alpha} =
-{}^{J\!}C_{+(p-\alpha+1)}$ for $\alpha = 2, \ldots, s-2$ and $\alpha
= s+1, \ldots, 2s-3$. An appropriate parametrization of the mapping
$c_-$ is given in Figure \ref{f:3}, where $C_{-0} = {}^{KJ\!} C_{-1}$,
$C_{-s} = -{}^{JK\!} C_{-(s-1)}$, and $C_{-\alpha} =
-{}^{J\!}C_{-(p-\alpha+1)}$ for $\alpha = 2, \ldots, s-2$ and $\alpha
= s+1, \ldots, 2s-3$.

Parameterize the mapping $\gamma$ in accordance with (\ref{e:12}),
where ${}^{K\!} \Gamma_1 = \Gamma_1^{-1}$, ${}^{K\!} \Gamma_s =
\Gamma_s^{-1}$, and  $\Gamma_\alpha = ({}^{J\!}
\Gamma_{p-\alpha+2})^{-1}$ for each $\alpha = 2, \ldots, s-1$ and
$\alpha = s+1, \ldots, 2s-2$. The Lie group $G_0$ is isomorphic to the
Lie group $\rmSp_{n_1}(\bbC) \times \rmGL_{n_2}(\bbC) \times \cdots
\times \rmGL_{n_{s-1}}(\bbC) \times \rmSp_{n_s}(\bbC)$, and the Toda
equation (\ref{e:5}) is equivalent to the system of equations
\begin{align*}
\partial_+ \left( \Gamma_1^{-1} \, \partial_- \Gamma_1^{} \right) &= -
\Gamma_1^{-1} C_{+1}^{} \, \Gamma_2^{} \, C_{-1}^{} +
{}^{K}(\Gamma_1^{-1}
C_{+1}^{} \, \Gamma_2^{} \, C_{-1}^{}), \\*
\partial_+ \left( \Gamma_2^{-1} \, \partial_- \Gamma_2^{} \right) &= -
\Gamma_2^{-1} C_{+2}^{} \, \Gamma_3^{} \, C_{-2}^{} + C_{-1}^{}
\Gamma_1^{-1} C_{+1}^{} \Gamma_2^{}, \\*
& \quad \vdots \\*
\partial_+ \left( \Gamma_{s-1}^{-1} \, \partial_- \Gamma_{s-1}^{}
\right)
&= - \Gamma_{s-1}^{-1} C_{+(s-1)}^{} \, \Gamma_s^{} \, C_{-(s-1)}^{} +
C_{-(s-2)}^{} \Gamma_{s-2}^{-1} C_{+(s-2)}^{} \Gamma_{s-1}^{}, \\*
\partial_+ \left( \Gamma_s^{-1} \, \partial_- \Gamma_s^{} \right) &= -
{}^{K}(C_{-(s-1)}^{} \Gamma_{s-1}^{-1} C_{+(s-1)}^{} \Gamma_s^{}) +
C_{-(s-1)}^{} \Gamma_{s-1}^{-1} C_{+(s-1)}^{} \Gamma_s^{}.
\end{align*}

Consider now the case of an odd $p = 2s - 1$, $s \ge 2$. Parameterize
the mapping $c_+$ as is given in Figure \ref{f:2}, where $C_{+0} =
{}^{JK\!} C_{+1}$, $C_{+s} = {}^{J\!} C_{+s}$, and $C_{+\alpha} =
-{}^{J\!}C_{+(p-\alpha+1)}$ for $\alpha = 2, \ldots, s-1$ and $\alpha
= s+1, \ldots, 2s-2$. An appropriate parametrization of the mapping
$c_-$ is given in Figure \ref{f:3}, where $C_{-0} = {}^{KJ\!} C_{-1}$,
$C_{-s} = {}^{J\!} C_{-s}$, and $C_{-\alpha} =
-{}^{J\!}C_{-(p-\alpha+1)}$ for $\alpha = 2, \ldots, s-1$ and $\alpha
= s+1, \ldots, 2s-2$. The mapping $\gamma$ is parameterized in
accordance with (\ref{e:12}), where ${}^{K\!} \Gamma_1 =
\Gamma_1^{-1}$, and  $\Gamma_\alpha = ({}^{J\!}
\Gamma_{p-\alpha+2})^{-1}$ for each $\alpha = 2, \ldots, 2s-1$. The
Lie group $G_0$ is isomorphic to the Lie group $\rmSp_{n_1}(\bbC)
\times \rmGL_{n_2}(\bbC) \times \cdots \times \rmGL_{n_s}(\bbC)$ and
the Toda equation (\ref{e:5}) is equivalent to the system of equations
\begin{align}
\partial_+ \left( \Gamma_1^{-1} \, \partial_- \Gamma_1^{} \right) &= -
\Gamma_1^{-1} C_{+1}^{} \, \Gamma_2^{} \, C_{-1}^{} +
{}^{K}(\Gamma_1^{-1}
C_{+1}^{} \, \Gamma_2^{} \, C_{-1}^{}), \notag \\*
\partial_+ \left( \Gamma_2^{-1} \, \partial_- \Gamma_2^{} \right) &= -
\Gamma_2^{-1} C_{+2}^{} \, \Gamma_3^{} \, C_{-2}^{} + C_{-1}^{}
\Gamma_1^{-1} C_{+1}^{} \Gamma_2^{}, \notag \\*
& \quad \vdots \label{e:34} \\*
\partial_+ \left( \Gamma_{s-1}^{-1} \, \partial_- \Gamma_{s-1}^{}
\right)
&= - \Gamma_{s-1}^{-1} C_{+(s-1)}^{} \, \Gamma_s^{} \, C_{-(s-1)}^{} +
C_{-(s-2)}^{} \Gamma_{s-2}^{-1} C_{+(s-2)}^{} \Gamma_{s-1}^{}, \notag
\\*
\partial_+ \left( \Gamma_s^{-1} \, \partial_- \Gamma_s^{} \right) &= -
\Gamma_s^{-1} C_{+s}^{} ({}^{J\!} \Gamma_s^{-1}) C_{-s}^{} +
C_{-(s-1)}^{} \Gamma_{s-1}^{-1} C_{+(s-1)}^{} \Gamma_s^{}. \notag
\end{align}
Remind that in the case under consideration ${}^{J\!} C_{+s} = C_{+s}$
and ${}^{J\!} C_{-s} = C_{-s}$.

\subsubsection{\mathversion{bold}$m_1 < M$}

In accordance with the results of Section \ref{s:4.1.2}, a
$\bbZ_M$-gradation of the Lie algebra $\gothsp_n(\bbC)$ of inner type
for which $\nu = 1$ and $m_1 < M$ is specified by a choice of positive
integers $n_\alpha$ satisfying the equality $\sum_{\alpha=1}^p
n_\alpha = n$ and the relations (\ref{e:26}), and positive integers
$k_\alpha$ such that $M - \sum_{\beta=1}^{p-1} k_\beta$ is an even
positive integer and which satisfy the relations (\ref{e:29}).

In the case of an odd $p = 2s - 1$, $s \ge 2$, we parameterize the
mapping $c_+$ as is given in Figure \ref{f:2}, where $C_{+0} =
{}^{J\!} C_{+0}$, $C_{+s} = -{}^{KJ\!} C_{+(s-1)}$ and $C_{+\alpha} =
-{}^{J\!}C_{+(p - \alpha)}$ for $\alpha = 1, \ldots, s-2$ and $\alpha
= s+1, \ldots, 2s - 2$. An appropriate parametrization of the mapping
$c_-$ is described by Figure~\ref{f:3}, where $C_{-0} = {}^{J\!}
C_{-0}$, $C_{-s} = -{}^{JK\!} C_{-(s-1)}$ and $C_{-\alpha} =
-{}^{J\!}C_{-(p - \alpha)}$ for $\alpha = 1, \ldots, s-2$ and $\alpha
= s + 1, \ldots, 2s - 2$. The mapping $\gamma$ can be parameterized in
accordance with (\ref{e:12}), where ${}^{K\!} \Gamma_s =
\Gamma_s^{-1}$, and  $\Gamma_\alpha = ({}^{J\!}
\Gamma_{p-\alpha+1})^{-1}$ for $\alpha = 1, \ldots, s-1$ and $\alpha =
s+1, \ldots, 2s-1$. The Lie group $G_0$ is isomorphic to
$\rmGL_{n_1}(\bbC) \times \cdots \times \rmGL_{n_{s-1}}(\bbC) \times
\rmSp_{n_s}(\bbC)$. One can show that in the case under consideration
the Toda equation (\ref{e:5}) is equivalent to the system
\begin{align*}
\partial_+ \left( \Gamma_1^{-1} \, \partial_- \Gamma_1^{} \right) &= -
\Gamma_1^{-1} C_{+1}^{} \, \Gamma_2^{} \, C_{-1}^{} + C_{-0}^{} \,
{}^{J\!} \Gamma_1 \, C_{+0}^{} \Gamma_1^{}, \\*
\partial_+ \left( \Gamma_2^{-1} \, \partial_- \Gamma_2^{} \right) &= -
\Gamma_2^{-1} C_{+2}^{} \, \Gamma_3^{} \, C_{-2}^{} + C_{-1}^{}
\Gamma_1^{-1} C_{+1}^{} \Gamma_2^{}, \notag \\*
& \quad \vdots \\*
\partial_+ \left( \Gamma_{s-1}^{-1} \, \partial_- \Gamma_{s-1}^{}
\right) &= - \Gamma_{s-1}^{-1} C_{+(s-1)}^{} \, \Gamma_s^{} \,
C_{-(s-1)}^{} + C_{-(s-2)}^{} \Gamma_{s-2}^{-1} C_{+(s-2)}^{}
\Gamma_{s-1}^{}, \\*
\partial_+ \left( \Gamma_s^{-1} \, \partial_- \Gamma_s^{} \right) &= -
{}^{K\!} (C_{-(s-1)}^{} \Gamma_{s-1}^{-1} C_{+(s-1)}^{} \Gamma_s^{}) +
C_{-(s-1)}^{} \Gamma_{s-1}^{-1} C_{+(s-1)}^{} \Gamma_s^{}.
\end{align*}
It is not difficult to get convinced that the substitutions
\[
\Gamma_\alpha \to {}^{J\!}(\Gamma_{s-\alpha+1}^{-1}), \quad \alpha =
1, \ldots, s-1, \qquad \Gamma_s \to {}^{K\!}(\Gamma_1^{-1})
\]
together with
\begin{gather*}
C_{\pm 0} \to {}^{J\!} C_{\pm s}, \qquad C_{+(s-1)} \to {}^{JK\!}
C_{+1}, \qquad C_{-(s-1)} \to {}^{KJ\!} C_{-1}, \\
C_{\pm \alpha} \to - {}^{J\!} C_{\pm (s-\alpha)}, \qquad \alpha = 1,
\ldots, s-2,
\end{gather*}
transform this system into the system (\ref{e:34}).

In the case of an even $p = 2s$ we come to the system (\ref{e:32})
where ${}^{J\!} C_{+0} = C_{+0}$, ${}^{J\!} C_{+s} = C_{+s}$,
${}^{J\!} C_{-0} = C_{-0}$, and ${}^{J\!} C_{-s} = C_{-s}$.

\subsection{\mathversion{bold}$\nu = -1$}

Following the consideration given in Section \ref{s:4.2}, we conclude
that a $\bbZ_M$-gradation of the Lie algebra $\gothsp_n(\bbC)$ of
inner type for which $\nu = -1$ is specified by a choice of positive
integers $n_\alpha$ satisfying the equality $\sum_{\alpha=1}^p
n_\alpha = n$ and the relations (\ref{e:26}), and positive integers
$k_\alpha$ such that $M - \sum_{\beta=1}^{p-1} k_\beta + 1$ is an even
positive integer, and which satisfy the relations (\ref{e:29}). It is
not difficult to see that using $\bbZ_M$-gradations of the type under
consideration we do not obtain new equations.

\section{Toda equations associated with loop groups of complex general
linear groups. Gradations of outer type} \label{s:6}

Let $a$ and $A$ denote an outer automorphism of the Lie group
$\rmGL_n(\bbC)$ of order $M$ and the corresponding outer automorphism
of the Lie algebra $\gothgl_n(\bbC)$ respectively. Consider the
$\bbZ_M$-gradation of the Lie algebra $\gothgl_n(\bbC)$ generated by
the automorphism $A$. Without any loss of generality we assume that
the automorphism $A$ is given by the relation
\begin{equation}
A(x) = - h \, {}^{J\!} x \, h^{-1}, \label{e:35}
\end{equation}
where $h$ is an element of $\rmGL_n(\bbC) \cap \rmD_n(\bbC)$, such
that
\begin{equation}
{}^{J\!}h \, h = I, \label{e:36}
\end{equation}
see Appendix \ref{a:c2}. From the equality (\ref{e:36}) it follows
that either $\det h = 1$, or $\det h = -1$. Since $h$ is a diagonal
matrix and satisfies the equality (\ref{e:36}), we conclude that the
equality $\det h = -1$ could be valid only if $n$ is odd. In this
case, multiplying $h$ by $-1$, we obtain a matrix with unit
determinant, which satisfies (\ref{e:36}) and generates the same
automorphism of $\gothgl_n(\bbC)$ as the initial matrix $h$. Thus, we
can assume that the automorphism $A$ generating the $\bbZ_M$-gradation
under consideration is given by the equality (\ref{e:35}), where $h$
is an element of the group $\rmSO_n(\bbC) \cap \rmD_n(\bbC)$.

Suppose now that $M$ is odd. The equality $A^M =
\id_{\gothgl_n(\bbC)}$ implies that
$- h^M \, {}^{J\!}x \, h^{-M} = x$ for any $x \in \gothgl_n(\bbC)$. It
is not difficult to show that it is impossible. Hence, $M$ cannot be
odd, and we assume that $M$ is even and equals $2N$. In this case we
obtain $h^M \, x \, h^{-M} = x$ for any $x \in \gothgl_n(\bbC)$.
Hence, the matrix $h$ satisfies the equality
\[
h^M = \nu I
\]
for some complex number $\nu$. It follows from this equality that
\[
{}^{J\!} (h^M) \, h^M = \nu^2 I.
\]
{}From the other hand, using (\ref{e:36}), we obtain
\[
{}^{J\!} (h^M) \, h^M = ({}^{J\!} h \, h)^M = I.
\]
Thus, the number $\nu$ should satisfy the equality $\nu^2 = 1$. This
means that either $\nu = 1$, or $\nu = -1$.

In the simplest case $h = I$. In this case there are only two
nontrivial grading subspaces,
\[
\gothg_{[0]_{2N}} = \{ x \in \gothgl_n(\bbC) \mid {}^{J\!} x = -x \}
\]
and
\[
\gothg_{[N]_{2N}} = \{ x \in \gothgl_n(\bbC) \mid {}^{J\!} x = x \}.
\]
The Lie group $G_0$ coincides with $\rmSO_n(\bbC)$ and we come to
Eq.~(\ref{e:5a}), where $\Gamma$ is a mapping from $\cal M$ to
$\rmSO_n(\bbC)$, the mappings $C_+$ and $C_-$ are mappings from
$\calM$ to the space of $n \times n$ complex matrices $x$ satisfying
the equality ${}^{J\!} x = x$.

Before proceeding to the consideration of a general case, it is useful
to make a few remarks.

Let $B$ be an arbitrary nonsingular matrix. For any $h \in
\rmGL_n(\bbC)$ the mapping $A: \gothgl_n(\bbC) \to \gothgl_n(\bbC)$
defined by the equality
\begin{equation}
A(x) = - h \, {}^{B\!} x \, h^{-1} \label{e:37}
\end{equation}
is an automorphism of $\gothgl_n(\bbC)$. By an inner automorphism of
$\gothgl_n(\bbC)$ generated by an element $k \in \rmGL_n(\bbC)$ the
automorphism $A$ is conjugated to an automorphism $A'$ of
$\gothgl_n(\bbC)$ defined by the equality
\[
A'(x) = - h' \, {}^{B'\!\!} x \, h^{\prime -1},
\]
where
\[
h' = k \, h \, k^{-1}, \qquad B' = {}^{t\!} k^{-1} B \, k^{-1}.
\]
Here if $B$ is a skew-diagonal symmetric matrix, then using an inner
automorphism generated by a diagonal matrix $k$ we can make $B$
coincide with the matrix~$J$. If $B$ is a skew-diagonal skew-symmetric
matrix, then using an inner automorphism also generated by a diagonal
matrix $k$ we can make $B$ coincide with the matrix~$K$.

Note also that for any $k \in \rmGL_n(\bbC)$ we can represent the
action of the automorphism $A$ given by (\ref{e:37}) on an element $x
\in \gothgl_n(\bbC)$ as
\[
A(x) = - h' \, {}^{B'\!\!} x \, h^{\prime -1}
\]
where
\[
h' = h \, k, \qquad B' =  B \, k.
\]

\subsection{\mathversion{bold}$\nu = 1$}

Represent each diagonal matrix element of the diagonal matrix $h$
either as $\epsilon_{M}^m$, or as $-\epsilon_{M}^m$, where $m$ is an
integer satisfying the condition $0 < m \le M/2 = N$. Let $m_\alpha$,
$\alpha = 1, \ldots, p$, be the different values of $m$ needed for
such representation taken in the decreasing order, $N \ge m_1 > m_2 >
\ldots > m_p > 0$. Denote by $n'_\alpha$ and $n''_\alpha$ the numbers
of matrix elements equal to $\epsilon_{M}^{m_\alpha}$ and
$-\epsilon_{M}^{m_\alpha}$, respectively, and by $n_\alpha$ the sum of
$n'_\alpha$ and $n''_\alpha$. Note that the integer $n - n_1$ is
always even.

\subsubsection{\mathversion{bold}$m_1 = N$}

The conditions $\det h = 1$ and ${}^{J\!} h \, h = I$ imply that the
integer $n'_1$ is even. Moreover, one can show that
\[
m_\alpha = N - m_{p-\alpha+2}, \qquad \alpha = 2, \ldots, p,
\]
and
\[
n'_\alpha = n''_{p-\alpha+2}, \qquad \alpha = 2, \ldots, p.
\]
The last equality, in particular, gives
\[
n_\alpha = n_{p-\alpha+2}, \qquad \alpha = 2, \ldots, p.
\]

Using automorphisms of $\gothgl_n(\bbC)$ interchanging simultaneously
rows and the corresponding columns of the matrices, one can bring the
element $h$ to the form
\[
\psset{xunit=2.7em, yunit=1.6em}
h = \left( \begin{pspicture}[.5](.3,3.4)(8.9,11.5)
\rput(1,11){$-I_{n'_1/2}$} \rput(2,10){$I_{n''_1}$}
\rput(3,9){$-I_{n'_1/2}$} \rput(4,8){$\epsilon_{M}^{m_2} I_{n'_2}$}
\rput(5,7){$-\epsilon_{M}^{m_2} I_{n''_2}$} \psddots(6,6)
\rput(7,5){$\epsilon_{M}^{m_{p}} I_{n'_{p}}$}
\rput(8,4){$-\epsilon_{M}^{m_{p}} I_{n''_{p}}$}
\end{pspicture} \right).
\]
Here the matrix $J$ goes to the matrix $B$ defined as
\[
B = \left( \begin{array}{cc}
J_{n_1} & 0 \\
0 & J_{n-n_1}
\end{array} \right).
\]
Now multiplying $h$ and $B$ from the right by the appropriate diagonal
matrix we bring the matrix $h$ to the form
\begin{equation}
\psset{xunit=2.7em, yunit=1.6em}
h = \left( \begin{pspicture}[.45](.4,.5)(4.6,4.4)
\rput(1,4){$\epsilon_M^{m_1} I_{n_1}$} \rput(2,3){$\epsilon_M^{m_2}
I_{n_2}$} \psddots(3,2) \rput(4,1){$\epsilon_M^{m_p} I_{n_p}$}
\end{pspicture} \right). \label{e:38}
\end{equation}
Here the matrix $B$ takes the form
\[
B = \left( \begin{array}{cc}
B' & 0 \\
0 & B''
\end{array} \right),
\]
where $B'$ is an $n_1 \times n_1$ skew-diagonal symmetric matrix, and
$B''$ is an $(n-n_1) \times (n-n_1)$ skew-diagonal skew-symmetric
matrix. Using an inner automorphism of $\gothgl_n(\bbC)$ generated by
a diagonal matrix we bring the matrix $B$ to the form
\begin{equation}
B = \left( \begin{array}{cc}
J_{n_1} & 0 \\
0 & K_{n-n_1}
\end{array} \right). \label{e:39}
\end{equation}
Here the matrix $h$ remains unchanged.

Thus, in the case under consideration the outer automorphism $A$ of
$\gothgl_n(\bbC)$
defined by the relation (\ref{e:35}) is conjugated to the automorphism
$A$ defined by the relation (\ref{e:37}) with $h$ and $B$ given by the
equalities (\ref{e:38}) and (\ref{e:39}) respectively. We assume that
the automorphism $A$ under consideration has this form.

It can be easily verified that
\[
{}^{B\!}({}^{B\!} x) = \tilde I{}^{-1} x \tilde I,
\]
where
\[
\tilde I = \left( \begin{array}{cc}
I_{n_1} & \\
& -I_{n-n_1}
\end{array} \right).
\]
Now using the equality
\[
{}^{B\!}h \, h = \tilde I,
\]
one obtains
\[
A^2(x) = h^2 \, x \, h^{-2}.
\]
Let us denote
\begin{equation}
\psset{xunit=2.7em, yunit=1.6em}
\tilde h = h^2 = \left(
\begin{pspicture}[.45](.4,.5)(4.6,4.4)
\rput(1,4){$\epsilon_N^{m_1} I_{n_1}$} \rput(2,3){$\epsilon_N^{m_2}
I_{n_2}$} \psddots(3,2) \rput(4,1){$\epsilon_N^{m_p} I_{n_p}$}
\end{pspicture} \right), \label{e:40}
\end{equation}
and rewrite the above relation as
\begin{equation}
A^2(x) = \tilde h \, x \, \tilde h^{-1}. \label{e:41}
\end{equation}

Let $k$ be an integer such that $0 \le k < 2N$. Assume that an element
$x$ belongs to the grading subspace with the grading index $[k]_{2N}$
of $\bbZ_{2N}$-gradation of $\gothgl_n(\bbC)$ defined by the
automorphism $A$. By definition we have
\begin{equation}
A(x) = \epsilon_{2N}^k \, x, \label{e:42}
\end{equation}
therefore,
\begin{equation}
A^2(x) = \epsilon_N^k \, x. \label{e:43}
\end{equation}
Hence, the element $x$ belongs to the grading subspace with the
grading index $[k]_N$ of inner type $\bbZ_N$-gradation of
$\gothgl_n(\bbC)$ defined by the automorphism $A^2$. Using the block
matrix representation (\ref{e:10}) and the relations (\ref{e:41}),
(\ref{e:40}), we see that the block $x_{\alpha \beta}$ of $x$ is
different from zero only if
\[
\epsilon_N^{m_\alpha - m_\beta} = \epsilon_N^k.
\]
There are four variants for the restrictions on possible values of
$m_\alpha$ and $m_\beta$ arising from this equality. They are
described by Table \ref{t:1}.
\begin{table}[ht]
\[
\arraycolsep 1em
\begin{array}{c|cc}
 & 0 \le k < N & N \le k < 2N \\ \hline
\vrule width 0pt height 2.5ex \alpha \le \beta & m_\alpha - m_\beta =
k & m_\alpha - m_\beta = k - N \\
\vrule width 0pt height 2.5ex \alpha > \beta & m_\alpha - m_\beta = k
- N & m_\alpha - m_\beta = k - 2N
\end{array}
\]
\caption{The restrictions on possible values of $m_\alpha$ and
$m_\beta$ for a fixed value of $k$} \label{t:1}
\end{table}

Introduce positive integers $k_\alpha = m_\alpha - m_{\alpha+1}$,
$\alpha = 1, \ldots, p-1$. They satisfy the relation
\[
k_\alpha = k_{p-\alpha+1}, \qquad \alpha = 2, \ldots, p-1.
\]
Similarly as in section \ref{s:4.1.1} one can show that the numbers
$k_\alpha$ satisfy the condition
\begin{equation}
\sum_{\beta=1}^{p-1} k_\beta + k_1 = N, \label{e:51}
\end{equation}
and that the integers $m_\alpha$ are connected with the numbers
$k_\alpha$ by the relation
\[
m_{\alpha} = \sum_{\beta = \alpha}^{p-1} k_\beta + k_1.
\]
Now one can get convinced that the grading structure of the
$\bbZ_{2N}$-gradation under consideration can be depicted by the
scheme given in Figure \ref{f:4}.
\begin{figure}[htb]
\[
\left( \begin{array}{c|c|c|c|c}
[0]_{2N} & [k_1]_{2N} & [k_1 + k_2]_{2N} & \cdots & \displaystyle
[\sum_{\alpha=1}^{p-1} k_\alpha]_{2N} \\
{}[N]_{2N} & [k_1 + N]_{2N} & [k_1 + k_2 + N]_{2N} & \cdots &
\displaystyle [\sum_{\alpha=1}^{p-1} k_\alpha + N]_{2N} \\\hline
-[k_1]_{2N} & [0]_{2N} & [k_2]_{2N} & \cdots & \displaystyle
\sum_{\alpha=2}^{p-1}
[k_\alpha]_{2N} \\
-[k_1 + N]_{2N} & [N]_{2N} & [k_2 + N]_{2N} & \cdots & \displaystyle
[\sum_{\alpha=2}^{p-1} k_\alpha + N]_{2N} \\ \hline
-[k_1 + k_2]_{2N} & - [k_2]_{2N} & [0]_{2N} & \cdots & \displaystyle
[\sum_{\alpha = 3}^{p-1} k_\alpha]_{2N} \\
-[k_1 + k_2 + N]_{2N} & - [k_2 + N]_{2N} & [N]_{2N} & \cdots &
\displaystyle
[\sum_{\alpha = 3}^{p-1} k_\alpha + N]_{2N} \\ \hline
\vdots & \vdots & \vdots & \ddots & \vdots \\ \hline
\displaystyle - [\sum_{\alpha = 1}^{p-1} k_\alpha]_{2N} & -
\displaystyle
[\sum_{\alpha=2}^{p-1} k_\alpha]_{2N} & -\displaystyle
[\sum_{\alpha=3}^{p-1}
k_\alpha]_{2N} & \cdots & [0]_{2N} \\
{} \displaystyle -[\sum_{\alpha = 1}^{p-1} k_\alpha + N]_{2N} &
\displaystyle - [\sum_{\alpha=2}^{p-1} k_\alpha + N]_{2N} &
\displaystyle - [\sum_{\alpha=3}^{p-1}
k_\alpha + N]_{2N} & \cdots & [N]_{2N}
\end{array} \right)
\]
\caption{The structure of an outer $\bbZ_{2N}$-gradation of
$\gothgl_n(\bbC)$} \label{f:4}
\end{figure}
Here the pairs of elements of the ring $\bbZ_{2N}$ are the possible
grading indices of the corresponding blocks in the block matrix
representation (\ref{e:10}) of a general element of $\gothgl_n(\mathbb
C)$. Note that this structure is obtained by using only the equality
(\ref{e:43}) which follows from the equality (\ref{e:42}). The latter
equality imposes additional restrictions.

Using again the block matrix representation (\ref{e:10}), we see that
the equality (\ref{e:42}) gives
\[
\epsilon^{m_\alpha - m_\beta}_{2N} \, ({}^{B\!} x)_{\alpha \beta} = -
\epsilon^k_{2N} \, x_{\alpha \beta}.
\]
Having in mind Table \ref{t:1}, we come to the restrictions on the
block structure of an element $x$ belonging to a grading subspace of
the $\bbZ_{2N}$-gradation under consideration collected in Table
\ref{t:2}.
\begin{table}[ht]
\[
\arraycolsep 1em
\begin{array}{c|cc}
 & 0 \le k < N & N \le k < 2N \\ \hline
\vrule width 0pt height 3ex \alpha \le \beta & x_{\alpha \beta} =
-({}^{B\!} x)_{\alpha \beta} & x_{\alpha \beta} = ({}^{B\!} x)_{\alpha
\beta} \\
\vrule width 0pt height 2.5ex \alpha > \beta & x_{\alpha \beta} =
({}^{B\!} x)_{\alpha \beta} & x_{\alpha \beta} = -({}^{B\!} x)_{\alpha
\beta}
\end{array}
\]
\caption{The restrictions on the structure of the blocks $x_{\alpha
\beta}$} \label{t:2}
\end{table}

Consider now the corresponding Toda equations. As in the cases
considered before, to obtain equations which can not be reduced to
Toda equations associated with finite dimensional group we should put
$k_\alpha = L$, $\alpha = 1, \ldots, p-1$. It follows from
(\ref{e:51}) that in this case we should assume that $N = p L$.

For the case of an even $p=2s-2$, $s \ge 2$, the group $G_0$ is
isomorphic to $\rmSO_{n_1}(\bbC) \times \rmGL_{n_2}(\bbC) \times
\cdots \times \rmGL_{n_{s-1}}(\bbC) \times \rmSp_{n_s}(\bbC)$. The
Toda equation (\ref{e:5}) is equivalent to the system
\begin{align*}
\partial_+ \left( \Gamma_1^{-1} \, \partial_- \Gamma_1^{} \right) &= -
\Gamma_1^{-1} C_{+1}^{} \, \Gamma_2^{} \, C_{-1}^{} + {}^{J\!}
(\Gamma_1^{-1}
C_{+1}^{} \, \Gamma_2^{} \, C_{-1}^{}), \\*
\partial_+ \left( \Gamma_2^{-1} \, \partial_- \Gamma_2^{} \right) &= -
\Gamma_2^{-1} C_{+2}^{} \, \Gamma_3^{} \, C_{-2}^{} + C_{-1}^{}
\Gamma_1^{-1} C_{+1}^{} \Gamma_2^{}, \\*
& \quad \vdots \\*
\partial_+ \left( \Gamma_{s-1}^{-1} \, \partial_- \Gamma_{s-1}^{}
\right)
&= - \Gamma_{s-1}^{-1} C_{+(s-1)}^{} \, \Gamma_s^{} \, C_{-(s-1)}^{} +
C_{-(s-2)}^{} \Gamma_{s-2}^{-1} C_{+(s-2)}^{} \Gamma_{s-1}^{}, \\*
\partial_+ \left( \Gamma_s^{-1} \, \partial_- \Gamma_s^{} \right) &= -
{}^{K\!} (C_{-(s-1)}^{} \Gamma_{s-1}^{-1} C_{+(s-1)}^{} \Gamma_s^{}) +
C_{-(s-1)}^{} \Gamma_{s-1}^{-1} C_{+(s-1)}^{} \Gamma_s^{},
\end{align*}
where ${}^{J\!} \Gamma_1 = \Gamma_1^{-1}$ and ${}^{K\!} \Gamma_s =
\Gamma_s^{-1}$. One can get convinced that using appropriate
substitutions it is possible to reduce the above system to the system
of the same form but with $J$ and $K$ interchanged.

For the case of an odd $p=2s-1$, $s \ge 2$, the Lie group $G_0$ is
isomorphic to the group
$\rmSO_{n_1}(\bbC) \times \rmGL_{n_2}(\bbC) \times \cdots \times
\rmGL_{n_s}(\bbC)$. The Toda equation (\ref{e:5}) is equivalent to the
system of equations (\ref{e:25}), where ${}^{J\!} \Gamma_1 =
\Gamma_1^{-1}$, ${}^{J\! } C_{+s} = C_{+s}$, and ${}^{J\! } C_{-s} =
C_{-s}$.

\subsubsection{\mathversion{bold}$m_1 < N$}

For the case when $\nu = 1$ and $m_1 < N$ the automorphism $A$ under
consideration is conjugated to the automorphism $A$ given by the
equality (\ref{e:37}), where $B = K$, the matrix $h$ has the form
(\ref{e:38}) and satisfies the equality ${}^{J\!} h \, h = I$. The
integers $k_\alpha$, $\alpha = 1, \ldots, p-1$, satisfy the relation
(\ref{e:29}). The structure of the $\bbZ_{2N}$-gradation generated by
the automorphism $A$ is described by Figure \ref{f:4}, and the
nontrivial blocks of an element $x$ belonging to the grading subspace
with the grading index $[k]_{2N}$ are subjected to the conditions
given in Table \ref{t:2}.

For the case of an odd $p=2s-1$ the Lie group $G_0$ is isomorphic to
the Lie group $\rmGL_{n_1}(\bbC) \times \cdots \times
\rmGL_{n_{s-1}}(\bbC) \times \rmSp_{n_s}(\bbC)$ and the Toda equation
(\ref{e:5}) is equivalent to a system of equations which can be
reduced to the system (\ref{e:34}), where ${}^{K\!} \Gamma_s =
\Gamma_s^{-1}$, and ${}^{J\!}C_{+0} = - C_{+0}$, ${}^{J\!}C_{-0} = -
C_{-0}$.

For the case of an even $p = 2s$ the Lie group $G_0$ is isomorphic to
the Lie group $\rmGL_{n_1}(\bbC) \times \cdots \times
\rmGL_{n_s}(\bbC)$ and the Toda equation (\ref{e:5}) is equivalent to
the system (\ref{e:32}), where ${}^{J\!} C_{+0} = -C_{+0}$, ${}^{J\!}
C_{+s} = C_{+s}$, ${}^{J\!} C_{-0} = -C_{-0}$, and ${}^{J\!} C_{-s} =
C_{-s}$.

\subsection{\mathversion{bold}$\nu = -1$}

In the case when $\nu = -1$, as for the inner $\bbZ_M$-gradations of
the complex orthogonal algebras and the complex symplectic algebras, we do
not obtain new equations.

\section{Toda equations associated with loop groups of complex
orthogonal groups. Gradations of outer type} \label{s:outerso}

Recall that the Lie group $\rmSO_n(\bbC)$ has outer automorphisms only
if $n$ is even. Therefore, we assume that $n$ is even and discard the
trivial case $n = 2$. Let $a$ be an outer automorphism of
$\rmSO_n(\bbC)$ of order $M$ and $A$ be the corresponding automorphism
of $\gothso_n(\bbC)$. Without any loss of generality we assume that
the automorphism $A$ is given by the equality
\[
A(x) = (h' u) \, x \, (h' u)^{-1},
\]
where the matrix $u$ is given by the equality (\ref{e:49}) and $h'$ is
an element of $\rmSO_n(\bbC) \cap \rmD_n(\bbC)$ satisfying the
relation $u \, h' \, u^{-1} = h'$, see Appendix \ref{a:c3}.

It is not difficult to show that the two central diagonal matrix
elements of the matrix $h'$ are equal either to $1$ or to $-1$. We
exclude the latter case multiplying the matrix $h'$ by $-1$. In other
words, we assume that the matrix $h = h' u$ has the form
\[
h = \left(\begin{array}{cccc}
h'' \\
& 0 & 1 \\
& 1 & 0 \\
& & & {}^{J\!} (h^{\prime \prime -1})
\end{array} \right),
\]
where $h''$ is an $(n-2)/2 \times (n-2)/2$ nonsingular diagonal
matrix. From the equality $A^M = \id_{\gothso_n(\bbC)}$ it follows
that
\[
h^M = \nu I
\]
for some complex number $\nu$. Using the above explicit form of the
matrix $h$, we conclude that it is possible only for an even $M = 2N$
and $\nu = 1$.

The conjugation by the matrix
\[
k = \frac{1}{\sqrt{2}} \left(\begin{array}{crcc}
\sqrt{2} \, I_{(n-2)/2} \\
& 1 & 1 \\
& -\rmi & \rmi \\
& & & \sqrt{2} \, I_{(n-2)/2}
\end{array} \right)
\]
brings the matrix $h$ to the form
\[
h = \left(\begin{array}{cccc}
h'' \\
& 1 & \\
& & -1 \\
& & & {}^{J\!} (h^{\prime \prime-1})
\end{array} \right),
\]
and maps $\gothso_n(\bbC)$ isomorphically onto the Lie algebra
$\gothgl_n^B(\bbC)$, where
\[
B = \left(\begin{array}{cccc}
& & & J_{(n-2)/2} \\
& 1 & \\
& & 1 \\
J_{(n-2)/2}
\end{array} \right).
\]
The same conjugation maps $\rmSO_n(\bbC)$ isomorphically onto the Lie
group $\rmGL_n^B(\bbC)$.

Note that $h$ is a diagonal matrix with the diagonal matrix elements
of the form $\epsilon_{2N}^m$, where we assume that $m$ is a
nonnegative integer satisfying the condition \hbox{$0 < m \le 2N$}.
Let $m_\alpha$, $\alpha = 1, \ldots, p$, be the different values of
$m$ taken in the decreasing order, $2N \ge m_1 > m_2 > \ldots > m_p >
0$, and $n_\alpha$, $\alpha = 1, \ldots, p$, be their multiplicities.
It is clear that $m_1 = 2N$. One can get convinced that $p$ is an even
number, let it be equal to $2s-2$. The integers $m_\alpha$ satisfy the
relations
\[
m_\alpha = 2 N - m_{p-\alpha+2}.
\]
In particular, one has $m_s = N$. It is also almost evident that $n_1$
and $n_s$ are odd.

Simultaneously permuting the rows and columns of the matrices, we
bring $h$ to the form (\ref{e:9}) with $\rho = 1$ and $M=2N$. Here the
Lie algebra under consideration is mapped isomorphically onto the Lie
algebra $\gothgl_n^B(\bbC)$ with $B$ given by the equality
(\ref{e:17}) and the Lie group under consideration is mapped onto Lie
group $\rmGL^B_n(\bbC)$ with the same $B$.

It is clear that the Lie group $G_0$ is isomorphic to the Lie group
$\rmSO_{n_1}(\bbC) \times \rmGL_{n_2}(\bbC) \times \cdots \times
\rmGL_{n_{s-1}}(\bbC) \times \rmSO_{n_s}(\bbC)$. The Toda equation
(\ref{e:5}) is equivalent now to the system (\ref{e:24}), where
${}^{J\!} \Gamma_1 = \Gamma_1^{-1}$ and ${}^{J\!} \Gamma_s =
\Gamma_s^{-1}$. Recall that in the case under consideration the
integers $n_1$ and $n_s$ are odd.

\section{Conclusions}

Thus, there are four types of Toda equations associated with loop
groups of complex classical Lie groups.

Let $n_\alpha$, $\alpha = 1, \ldots, p$, be a set of positive
integers. A Toda equation of the first type is equivalent to the
system of equations
\begin{align*}
\partial_+ \left( \Gamma_1^{-1} \, \partial_- \Gamma_1^{} \right) &= -
\Gamma_1^{-1} C_{+1}^{} \, \Gamma_2^{} \, C_{-1}^{} + C_{-0}^{}
\Gamma_p^{-1} C_{+0}^{} \Gamma_1^{}, \\*
\partial_+ \left( \Gamma_2^{-1} \, \partial_- \Gamma_2^{} \right) &= -
\Gamma_2^{-1} C_{+2}^{} \, \Gamma_3^{} \, C_{-2}^{} + C_{-1}^{}
\Gamma_1^{-1} C_{+1}^{} \Gamma_2^{}, \\*
& \quad \vdots \\*
\partial_+ \left( \Gamma_{p-1}^{-1} \, \partial_- \Gamma_{p-1}^{}
\right)
&= - \Gamma_{p-1}^{-1} C_{+(p-1)}^{} \, \Gamma_p^{} \, C_{-(p-1)}^{} +
C_{-(p-2)}^{} \Gamma_{p-2}^{-1} C_{+(p-2)}^{} \Gamma_{p-1}^{}, \\*
\partial_+ \left( \Gamma_p^{-1} \, \partial_- \Gamma_p^{} \right) &= -
\Gamma_p^{-1} C_{+p}^{} \, \Gamma_1^{} \, C_{-p}^{} + C_{-(p-1)}^{}
\Gamma_{p-1}^{-1} C_{+(p-1)}^{} \Gamma_p^{}.
\end{align*}
Here $\Gamma_\alpha$ for each $\alpha = 1,\ldots, p$ is a mapping from
the manifold of independent variables $\calM$ to the Lie group
$\rmGL_{n_\alpha}(\bbC)$. $C_{+\alpha}$
and $C_{-\alpha}$ for each $\alpha = 1, \ldots, p-1$ are fixed
mappings from $\calM$ to the space of $n_\alpha \times n_{\alpha+1}$
and $n_{\alpha+1} \times n_\alpha$ complex matrices respectively.
$C_{+0}$ and $C_{-0}$ are mappings from $\calM$ to the space of $n_p
\times n_1$ and $n_1 \times n_p$ complex matrices respectively. The
mappings $C_{-\alpha}$ and $C_{+\alpha}$ satisfy the conditions
\[
\partial_+ C_{-\alpha} = 0, \qquad \partial_- C_{+\alpha} = 0.
\]
The mappings $C_{-\alpha}$ and $C_{+\alpha}$ for the other types of
Toda equations should also satisfy these conditions.

To describe the Toda equations of the other types we introduce a set
of positive integers $n_\alpha$, $\alpha = 1, \ldots, s$. A Toda
equation of the second type is equivalent to the system of equations
\begin{align*}
\partial_+ \left( \Gamma_1^{-1} \, \partial_- \Gamma_1^{} \right) &= -
\Gamma_1^{-1} C_{+1}^{} \, \Gamma_2^{} \, C_{-1}^{} + {}^{F_1\!}
(\Gamma_1^{-1}
C_{+1}^{} \, \Gamma_2^{} \, C_{-1}^{}), \\*
\partial_+ \left( \Gamma_2^{-1} \, \partial_- \Gamma_2^{} \right) &= -
\Gamma_2^{-1} C_{+2}^{} \, \Gamma_3^{} \, C_{-2}^{} + C_{-1}^{}
\Gamma_1^{-1} C_{+1}^{} \Gamma_2^{}, \\*
& \quad \vdots \\*
\partial_+ \left( \Gamma_{s-1}^{-1} \, \partial_- \Gamma_{s-1}^{}
\right)
&= - \Gamma_{s-1}^{-1} C_{+(s-1)}^{} \, \Gamma_s^{} \, C_{-(s-1)}^{} +
C_{-(s-2)}^{} \Gamma_{s-2}^{-1} C_{+(s-2)}^{} \Gamma_{s-1}^{}, \\*
\partial_+ \left( \Gamma_s^{-1} \, \partial_- \Gamma_s^{} \right) &= -
{}^{F_2\!} (C_{-(s-1)}^{} \Gamma_{s-1}^{-1} C_{+(s-1)}^{} \Gamma_s^{})
+
C_{-(s-1)}^{} \Gamma_{s-1}^{-1} C_{+(s-1)}^{} \Gamma_s^{}.
\end{align*}
Here each of $F_1$ and $F_2$ can be either $J$ or $K$. $\Gamma_1$ is a
mapping from $\calM$ either to the Lie group $\rmSO_{n_1}(\bbC)$ or to
the Lie group $\rmSp_{n_1}(\bbC)$, subject to the choice of $F_1$.
Similarly, $\Gamma_s$ is a mapping from $\calM$ either to the Lie
group $\rmSO_{n_s}(\bbC)$ or to the Lie group $\rmSp_{n_s}(\bbC)$,
subject to the choice of $F_2$.
For each $\alpha = 2,\ldots, s-1$ the mapping $\Gamma_\alpha$ is a
mapping from $\calM$ to the Lie group $\rmGL_{n_\alpha}(\bbC)$.
$C_{+\alpha}$ and $C_{-\alpha}$ for each $\alpha = 1, \ldots, s-1$ are
fixed mappings from $\calM$ to the space of $n_\alpha \times
n_{\alpha+1}$ and $n_{\alpha+1} \times n_\alpha$ complex matrices
respectively.

A Toda equation of the third type is equivalent to the system of
equations
\begin{align*}
\partial_+ \left( \Gamma_1^{-1} \, \partial_- \Gamma_1^{} \right) &= -
\Gamma_1^{-1} C_{+1}^{} \, \Gamma_2^{} \, C_{-1}^{} + {}^{F\!}
(\Gamma_1^{-1} C_{+1}^{} \, \Gamma_2^{} \, C_{-1}^{}), \\*
\partial_+ \left( \Gamma_2^{-1} \, \partial_- \Gamma_2^{} \right) &= -
\Gamma_2^{-1} C_{+2}^{} \, \Gamma_3^{} \, C_{-2}^{} + C_{-1}^{}
\Gamma_1^{-1} C_{+1}^{} \Gamma_2^{}, \\*
& \quad \vdots \\*
\partial_+ \left( \Gamma_{s-1}^{-1} \, \partial_- \Gamma_{s-1}^{}
\right) &= - \Gamma_{s-1}^{-1} C_{+(s-1)}^{} \, \Gamma_s^{} \,
C_{-(s-1)}^{} + C_{-(s-2)}^{} \Gamma_{s-2}^{-1} C_{+(s-2)}^{}
\Gamma_{s-1}^{}, \\*
\partial_+ \left( \Gamma_s^{-1} \, \partial_- \Gamma_s^{} \right) &= -
\Gamma_s^{-1} C_{+s}^{} {}^{J\!} (\Gamma_s^{-1}) C_{-s}^{} +
C_{-(s-1)}^{} \Gamma_{s-1}^{-1} C_{+(s-1)}^{} \Gamma_s^{}.
\end{align*}
Here $F$ is either $J$ or $K$. $\Gamma_1$ is a mapping from $\calM$
either to the Lie group $\rmSO_{n_1}(\bbC)$ or to the Lie group
$\rmSp_{n_1}(\bbC)$, subject to the choice of $F$. For each $\alpha =
2,\ldots, s$ the mapping $\Gamma_\alpha$ is a mapping from $\calM$ to
the Lie group $\rmGL_{n_\alpha}(\bbC)$. $C_{+\alpha}$ and
$C_{-\alpha}$ for each $\alpha = 1, \ldots, s-1$ are fixed mappings
from $\calM$ to the space of $n_\alpha \times n_{\alpha+1}$ and
$n_{\alpha+1} \times n_\alpha$ complex matrices respectively. The
mappings $C_{+s}$ and $C_{-s}$ satisfy either the relations ${}^{J\!}
C_{+s} = - C_{+s}$ and ${}^{J\!} C_{-s} = - C_{-s}$ or the relations
${}^{J\!} C_{+s} = C_{+s}$ and ${}^{J\!} C_{-s} = C_{-s}$.

A Toda equation of the fourth type is equivalent to the system of
equations
\begin{align*}
\partial_+ \left( \Gamma_1^{-1} \, \partial_- \Gamma_1^{} \right) &= -
\Gamma_1^{-1} C_{+1}^{} \, \Gamma_2^{} \, C_{-1}^{} + C_{-0}^{}
{}^{J\!} \Gamma_1 C_{+0}^{} \Gamma_1^{}, \\*
\partial_+ \left( \Gamma_2^{-1} \, \partial_- \Gamma_2^{} \right) &= -
\Gamma_2^{-1} C_{+2}^{} \, \Gamma_3^{} \, C_{-2}^{} + C_{-1}^{}
\Gamma_1^{-1} C_{+1}^{} \Gamma_2^{}, \\*
& \quad \vdots \\*
\partial_+ \left( \Gamma_{s-1}^{-1} \, \partial_- \Gamma_{s-1}^{}
\right) &= - \Gamma_{s-1}^{-1} C_{+(s-1)}^{} \, \Gamma_s^{} \,
C_{-(s-1)}^{} + C_{-(s-2)}^{} \Gamma_{s-2}^{-1} C_{+(s-2)}^{}
\Gamma_{s-1}^{}, \\*
\partial_+ \left( \Gamma_s^{-1} \, \partial_- \Gamma_s^{} \right) &= -
\Gamma_s^{-1} C_{+s}^{} {}^{J\!} (\Gamma_s^{-1}) C_{-s} +
C_{-(s-1)}^{} \Gamma_{s-1}^{-1} C_{+(s-1)}^{} \Gamma_s^{}.
\end{align*}
Here for each $\alpha = 1, \ldots, s$ the mapping $\Gamma_\alpha$ is a
mapping from $\calM$ to the Lie group $\rmGL_{n_\alpha}(\bbC)$.
$C_{+\alpha}$ and $C_{-\alpha}$ for each $\alpha = 1, \ldots, s-1$ are
fixed mappings from $\calM$ to the space of $n_\alpha \times
n_{\alpha+1}$ and $n_{\alpha+1} \times n_\alpha$ complex matrices.
$C_{+0}$ and $C_{-0}$ are fixed mappings from $\calM$ to the space of
$n_1 \times n_1$ complex matrices. The mappings $C_{+s}$ and $C_{-s}$
satisfy either the relations ${}^{J\!} C_{+s} = - C_{+s}$ and
${}^{J\!} C_{-s} = - C_{-s}$ or the relations ${}^{J\!} C_{+s} =
C_{+s}$ and ${}^{J\!} C_{-s} = C_{-s}$. The mappings $C_{+0}$ and
$C_{-0}$ satisfy either the relations ${}^{J\!} C_{+0} = - C_{+0}$ and
${}^{J\!} C_{-0} = - C_{-0}$ or the relations ${}^{J\!} C_{+0} =
C_{+0}$ and ${}^{J\!} C_{-0} = C_{-0}$.

\subsection*{Acknowledgements}

The work of A.V.R. was supported in part by the Russian Foundation for
Basic Research under grants \#04--01--00352 and \#07--01--00234. Kh.S.N.
was supported in part by the grant of the President of the
Russian Federation NS-7293.2006.2 (Government contract No. 02.445.11.7370),
by the Russian Foundation for
Basic Research under grant \#07--01--00234, and by the
Alexander von Humboldt-Stiftung, under a follow-up fellowship program;
he is grateful to Stefan Theisen for discussions and kind hospitality
at the Max-Planck-Institut f\"ur Gravitationsphysik --
Albert-Einstein-Institut in Potsdam.

\makeatletter
\@addtoreset{equation}{section}
\def\theequation{\thesection.\arabic{equation}}
\makeatother

\appendix

\section{Automorphisms of Lie groups and Lie algebras}

Let $G$ be a group. A bijective mapping $a \colon G \to G$ is called an
{\em automorphism\/} of $G$ if
\[
a(g_1 g_2) = a(g_1) a(g_2)
\]
for all $g_1, g_2 \in G$. With respect to the composition operation the
set $\Aut G$ of all automorphisms of the group $G$ is a group called the
{\em automorphism group\/} of $G$. When $G$ is a real (complex) Lie
group, an automorphism $a$ of $G$ is called a {\em Lie group
automorphism\/} if $a$ is a diffeomorphism (biholomorphic mapping).
Usually for the case of a Lie group, saying `automorphism' one means
`Lie group automorphism', and saying `automorphism group' one means the
group formed by all Lie group automorphisms, if the opposite is not
stated explicitly.

Let $\gothg$ be a Lie algebra. A nonsingular linear mapping $A \colon
\gothg \to \gothg$ is called an {\em automorphism\/} of $\gothg$ if
\[
A([x_1, x_2]) = [A(x_1), A(x_2)]
\]
for all $x_1, x_2 \in \gothg$. With respect to the composition operation
the set $\Aut \gothg$ of all automorphisms of $\gothg$ is a group called
the {\em automorphism group\/} of $\gothg$. The group $\Aut \gothg$ is a
Lie subgroup of the Lie group $\rmGL(\gothg)$.

Let $G$ be a Lie group. We identify the Lie algebra $\gothg$ of $G$ with
the tangent space $T_e G$, where $e$ is the identity element of $G$. Let
$a$ be an automorphism of $G$, the differential of the mapping $a$ at
$e$ is an automorphism of $\gothg$. The mapping $a \in \Aut G \mapsto
a_{*e} \in \Aut \gothg$ is a homomorphism. In the case when $G$ is
connected this mapping is an injective homomorphism, and when $G$ is
simply connected it is an isomorphism.

For any real (complex) connected Lie group $G$ the group $\Aut G$ can be
supplied in a natural way with the structure of a real (complex) Lie
group so that the action of $\Aut G$ on $G$ be smooth (holomorphic).

For an arbitrary element $h$ of a group $G$ the conjugation
$\rmC(g)\colon h \in G \mapsto g \, h \, g^{-1} \in G$ is an
automorphism of $G$. Such an automorphism is said to be an {\em inner
automorphism\/}. The corresponding automorphism $\rmC(g)_{*e}$ of
$\gothg$ is denoted $\Ad(g)$.
Any automorphism of $G$ which is not inner is called an {\em outer
automorphism\/} of $G$. The inner automorphisms of $G$ form a subgroup
$\Inn G$ of $\Aut G$ called the {\em inner automorphism group\/} of $G$.
When $G$ is a connected Lie group, the group $\Inn G$ is a Lie subgroup
of $\Aut G$.

For a general Lie group $G$ the group $\Inn G$ is a normal subgroup of
$\Aut G$ and one can consider the quotient group $\Out G = \Aut G/\Inn
G$. It is customary, slightly abusing language, to call this group the
{\em outer automorphism group\/} of $G$. When $G$ is a connected real
(complex) Lie group and the group $\Inn G$ is a closed subgroup of $\Aut
G$, the group $\Out G$ has the natural structure of a real (complex) Lie
group.

A linear mapping $D$ from a Lie algebra $\gothg$ to itself is called a
{\em derivation\/} of $\gothg$ if
\[
D([x_1, x_2]) = [D(x_1), x_2] + [x_1, D(x_2)]
\]
for all $x_1, x_2 \in \gothg$. A linear combination of two derivations
of $\gothg$ and their commutator are again derivations of $\gothg$.
Hence, the set of all derivations of $\gothg$ is a Lie algebra denoted
$\Der \gothg$. The Lie algebra of the Lie group $\Aut \gothg$ is
isomorphic to $\Der \gothg$ and usually identified with it.

For any $x \in \gothg$ the linear mapping $\ad(x): y \in \gothg \mapsto
[x, y] \in \gothg$ is a derivation of $\gothg$ called an {\em inner
derivation\/}. The set of all inner derivations of $\gothg$ is an ideal
of $\Der \gothg$. The corresponding connected Lie subgroup of the Lie
group $\Aut \gothg$ is called the {\em inner automorphism group\/} of
$\gothg$ and denoted $\Inn \gothg$. An element of $\Inn \gothg$ is
called an {\em inner automorphism\/} of $\gothg$. An element of $\Aut
\gothg$ which does not belong to $\Inn \gothg$ is said to be an {\em
outer automorphism\/} of $\gothg$. The quotient group $\Aut \gothg/\Inn
\gothg$ is called the {\em outer automorphism group\/} of $\gothg$.

Let $G$ be a connected Lie group. In this case the restriction of the
mapping $a \in \Aut G \mapsto a_{*e} \in \Aut \gothg$ to $\Inn G$ is an
isomorphism of the Lie groups $\Inn G$ and $\Inn \gothg$. If~$G$ is
simply connected, the groups $\Out G$ and $\Out \gothg$ are isomorphic.

\section{Automorphisms of complex simple Lie algebras}

Let $\gothg$ be a complex simple Lie algebra of rank $r$. It is well
known that any derivation of $\gothg$ is an inner derivation. Therefore,
the connected component of the identity of $\Aut \gothg$ coincides with
the group $\Inn \gothg$ which in this case is a closed Lie subgroup of
$\Aut \gothg$. Here the group $\Out \gothg$ is discrete.

Let $\gothh$ be a Cartan subalgebra of $\gothg$, $\Delta$ be the root
system of $\gothg$ with respect to $\gothh$, and $\Pi = \{\alpha_1,
\ldots, \alpha_r\}$ be a base of $\Delta$. Denote by $(\cdot, \cdot)$ a
bilinear form on $\gothh^*$ induced by the restriction of the Killing
form of $\gothg$ to $\gothh$. A permutation $\sigma \in S_r$ is called
an {\em automorphism\/} of the {\em base\/} $\Pi$ if
\[
\frac{(\alpha_{\sigma(i)}, \alpha_{\sigma(j)})}{(\alpha_{\sigma(j)},
\alpha_{\sigma(j)})} = \frac{(\alpha_i, \alpha_j)}{(\alpha_j,
\alpha_j)}
\]
for all $i,j = 1, \ldots, r$. All automorphisms of $\Pi$ form a group
called the {\em automorphism group of\/} the {\em base\/} $\Pi$ and
denoted $\Aut \Pi$. Visually an automorphism of $\Pi$ can be represented
as a permutation of the vertices of the corresponding Dynkin diagram
which leaves it invariant. Therefore, it is customary to call an
automorphism of $\Pi$ an {\em automorphism of\/} the {\em Dynkin
diagram\/} of $\Pi$ and the group $\Aut \Pi$ the {\em automorphism group
of\/} the {\em Dynkin diagram\/}.

Let $h_i$, $i = 1, \ldots, r$ be Cartan generators of $\gothh$, $x_{+i}$
and $x_{-i}$, $i = 1, \ldots, r$ be Chevalley generators of $\gothg$.
For any $\sigma \in \Aut \Pi$ there exists a unique automorphism
$\Sigma$ of $\gothg$ such that $\Sigma(h_i) = h_{\sigma(i)}$,
$\Sigma(x_{+i}) = x_{+\sigma(i)}$ and $\Sigma(x_{-i}) = x_{-\sigma(i)}$
for each $i = 1, \ldots, r$. Hence, one can identify the group $\Aut
\Pi$ with a subgroup of $\Aut \gothg$. It can be also shown that
\[
\Aut \gothg = \Inn \gothg \rtimes \Aut \Pi,
\]
see, for example, the monograph \cite{OniVin90}.

In this paper we are interested in finite order automorphisms of Lie
algebras up to conjugations, which can be lifted up to automorphisms of
the corresponding Lie groups. Here the following statement appears to be
very useful. Any finite order automorphism of $\gothg$ is conjugated by
an inner automorphism of $\gothg$ to an automorphism $A$ of the form
\[
A = \exp \ad (x) \circ \Sigma,
\]
where $\Sigma$ is the automorphism of $\gothg$ corresponding to some
automorphism $\sigma$ of $\Pi$, and $x$ is an element of $\gothh$ such
that $\Sigma(x) = x$, see, for example, \cite[Proposition 8.1]{Kac94}.

\section{Complex classical Lie groups and their finite order
automorphisms}

By complex classical Lie groups we mean the Lie groups $\rmGL_n(\bbC)$,
$\rmO_n(\bbC)$ and $\rmSp_{2n}(\bbC)$. Let us recall the definition of
these Lie groups and describe their automorphisms.

\subsection{Complex general linear groups}  \label{a:c2}

The {\em complex general linear group\/} $\rmGL_n(\bbC)$ is formed by
all nonsingular complex \hbox{$n \times n$} matrices with matrix
multiplication as the group law. It is a complex connected Lie group.
Identifying a complex \hbox{$n \times n$} matrix $g$ with a linear
endomorphism of the vector space $\bbC^n$, whose matrix with respect to
the standard basis of $\bbC$ coincides with $g$, we identify
$\rmGL_n(\bbC)$ with the group of linear automorphisms of $\bbC^n$. The
Lie algebra $\gothgl_n(\bbC)$ of the Lie group $\rmGL_n(\bbC)$ is formed
by all complex $n \times n$ matrices with matrix commutator as the Lie
algebra law.

The Lie group $\rmGL_n(\bbC)$ and the Lie algebra $\gothgl_n(\bbC)$ are
not simple. It has a normal subgroup formed by the matrices of the form
$ \kappa I$, where $\kappa \in \bbC^\times$.
Actually $\rmGL_n(\bbC)$ is isomorphic to the Lie group $(\bbC^\times \!
\times \rmSL_n(\bbC))/\bbZ_n$, where $\rmSL_n(\bbC)$ is the {\em complex
special linear group\/} formed by the elements of $\rmGL_n(\bbC)$ with
unit determinant. The Lie algebra $\gothgl_n(\bbC)$ is the direct sum of
the Lie subalgebra formed by the complex $n \times n$ matrices
proportional to the unit $n \times n$ matrix and the complex special
linear Lie algebra $\gothsl_n(\bbC)$ formed by the complex traceless $n
\times n$ matrices. The Lie group $\rmSL_n(\bbC)$ is connected and
simple. The Lie algebra $\gothsl_n(\bbC)$ is also simple.

The rank $r$ of the Lie algebra $\gothsl_n(\bbC)$ is equal to $n-1$. The
corresponding Dynkin diagram is of the form
\[
\psset{unit=1pt, linewidth=.5pt}%
\begin{pspicture}[.5](0,0)(90,11)
  \psline(2,2)(32,2)
  \pscircle*[linecolor=white](2,2){2.25}
  \pscircle(2,2){2.25}
  \pscircle*[linecolor=white](22,2){2.25}
  \pscircle(22,2){2.25}
  \pscircle*(38,2){1}
  \pscircle*(45,2){1}
  \pscircle*(52,2){1}
  \psline(58,2)(86,2)
  \pscircle*[linecolor=white](68,2){2.25}
  \pscircle(68,2){2.25}
  \pscircle*[linecolor=white](88,2){2.25}
  \pscircle(88,2){2.25}
  \rput[b](2,7){$\scriptstyle 1$}
  \rput[b](22,7){$\scriptstyle 2$}
  \rput[b](68,7){$\scriptstyle r-1$}
  \rput[b](88,7){$\scriptstyle r$}
\end{pspicture}
\]
It is clear that there is only one nontrivial automorphism of the Dynkin
diagram of $\gothsl_n(\bbC)$ given by the equality
\begin{equation}
\sigma(i) = r-i+1. \label{e:44}
\end{equation}
Denote by $\rmD_n(\bbC)$ the Lie group formed by all nonsingular complex
diagonal $n \times n$ matrices, and by $\gothd_n(\bbC)$ its Lie algebra.
Choose the Lie algebra $\gothsl_n(\bbC) \cap \gothd_n(\bbC)$ as a Cartan
subalgebra of $\gothsl_n(\bbC)$. In this case the Cartan and Chevalley
generators can be chosen as
\[
h_i = e_{ii} - e_{i+1,i+1}, \qquad x_{+i} = e_{i,i+1}, \qquad x_{-i} =
e_{i+1,i}.
\]
Here and below $e_{ij}$ is the matrix with the matrix elements
\[
(e_{ij})_{kl} = \delta_{ik} \delta_{jl}.
\]
One can verify that with this choice of Cartan and Chevalley generators
the automorphism $\Sigma$ defined as
\[
\Sigma(x) = - d \, {}^{J\!} x \, d^{-1},
\]
where $d$ is an element of $\rmSL_n(\bbC) \cap \rmD_n(\bbC)$ such that
\[
d \, x_{+i} \, d^{-1} = - x_{+i}, \qquad d \, x_{-i} \, d^{-1} = -
x_{-i}
\]
for each $i = 1, \ldots, r$, is an automorphism of $\gothsl_n(\bbC)$
induced by the automorphism $\sigma$ of the Dynkin diagram of
$\gothsl_n(\bbC)$ defined by the equality (\ref{e:44}). It is convenient
to assume that ${}^{J\!} d \, d = I$.

Thus, a finite order inner automorphism of the Lie algebra
$\gothsl_n(\bbC)$ is conjugated by an inner automorphism of
$\gothsl_n(\bbC)$ to an automorphism $A$ given by the equality
\begin{equation}
A(x) = h \, x \, h^{-1}, \label{e:45}
\end{equation}
where $h$ is an element of $\rmSL_n(\bbC) \cap \rmD_n(\bbC)$, and a
finite order outer automorphism of $\gothsl_n(\bbC)$ is conjugated by an
inner automorphism of $\gothsl_n(\bbC)$ to an automorphism $A$ given by
the equality
\[
A(x) = - h' d \, {}^{J\!} x \, \, d^{-1} h^{\prime -1},
\]
where $h'$ is an element of $\rmSL_n(\bbC) \cap \rmD_n(\bbC)$ satisfying
the relation
\[
d \, {}^{J\!}h' \, d^{-1} = h^{\prime -1}.
\]
Denoting $h = h' d$, we see that a finite order outer automorphism of
$\gothsl_n(\bbC)$ is conjugated by an inner automorphism of
$\gothsl_n(\bbC)$ to an automorphism $A$ given by the equality
\[
A(x) = - h \, {}^{J\!} x \, \, h^{-1},
\]
where $h$ is an element of $\rmSL_n(\bbC) \cap \rmD_n(\bbC)$ satisfying
the relation ${}^{J\!}h = h^{-1}$.

Now it is not difficult to understand that a finite order inner
automorphism of the Lie group $\rmSL_n(\bbC)$ is conjugated by an inner
automorphism of $\rmSL_n(\bbC)$ to an automorphism $a$ given by the
equality
\begin{equation}
a(g) = h \, g \, h^{-1}, \label{e:46}
\end{equation}
where $h$ is an element of $\rmSL_n(\bbC) \cap \rmD_n(\bbC)$, and a
finite order outer automorphism of $\rmSL_n(\bbC)$ is conjugated by an
inner automorphism of $\rmSL_n(\bbC)$ to an automorphism $a$ given by
the equality
\begin{equation}
a(g) = h \, {}^{J\!}(g^{-1}) \, h^{-1}, \label{e:47}
\end{equation}
where $h$ is an element of $\rmSL_n(\bbC) \cap \rmD_n(\bbC)$ satisfying
the relation ${}^{J\!} h = h^{-1}$.

One can easily extend the automorphisms of $\rmSL_n(\bbC)$ described by
the relations (\ref{e:46}) and (\ref{e:47}) to automorphisms of
$\rmGL_n(\bbC)$ respectively. It can be shown that any finite order
inner automorphism of $\rmGL_n(\bbC)$ is conjugated by an inner
automorphism of $\rmGL_n(\bbC)$ to an automorphism defined by the
relation (\ref{e:46}), and any finite order outer automorphism of
$\rmGL_n(\bbC)$ is conjugated by an inner automorphism of
$\rmGL_n(\bbC)$ to an automorphism defined by the relation
(\ref{e:47}).

The other complex classical Lie groups can be defined as Lie subgroups
of the Lie group $\rmGL_n(\bbC)$ via the following general procedure.
Let $B$ be a complex nonsingular $n \times n$ matrix. It is not
difficult to get convinced that the elements $g$ of $\rmGL_n(\bbC)$
satisfying the condition
\[
{}^{B\!} g = g^{-1}
\]
form a Lie subgroup of $\rmGL_n(\bbC)$ which we denote by
$\rmGL_n^B(\bbC)$. One can get convinced that the Lie algebra
$\gothgl_n^B(\bbC)$ of $\rmGL_n^B(\bbC)$ is a subalgebra of the Lie
algebra $\gothgl_n(\bbC)$ formed by the complex $n \times n$ matrices
$x$ satisfying the condition
\[
{}^{B\!} x = -x.
\]

\subsection{Complex orthogonal groups} \label{a:c3}

For any symmetric nonsingular $n \times n$ matrix $B$ the Lie group
$\rmGL_n^B(\bbC)$ is isomorphic to the Lie group $\rmGL_n^J(\bbC)$. This
group is called the {\em complex orthogonal group\/} and denoted
$\rmO_n(\bbC)$. For an element $g \in \rmO_n(\bbC)$ from the equality
${}^{J\!} g = g^{-1}$ we obtain that $\det g$ is equal either to $1$, or
to $-1$. The elements of $\rmO_n(\bbC)$ with unit determinant form a
connected Lie subgroup of $\rmO_n(\bbC)$ called the {\em complex special
orthogonal group\/} and denoted $\rmSO_n(\bbC)$. This subgroup is the
connected component of the identity of $\rmO_n(\bbC)$. The Lie algebra
of $\rmSO_n(\bbC)$ is denoted $\gothso_n(\bbC)$. It is clear that the
Lie algebra of $\rmO_n(\bbC)$ coincides with the Lie algebra of
$\rmSO_n(\bbC)$. The Lie group $\rmSO_n(\bbC)$ and the Lie algebra
$\gothso_n(\bbC)$ are simple.

For an odd $n=2r+1$ the rank of the Lie algebra $\gothso_n(\bbC)$ is
equal to $r$ and the corresponding Dynkin diagram has the form
\[
\psset{unit=1pt, linewidth=.5pt}%
\begin{pspicture}[.5](0,0)(90,13)
  \psline(2,4)(32,4)
  \pscircle*[linecolor=white](2,4){2.25}
  \pscircle(2,4){2.25}
  \pscircle*[linecolor=white](22,4){2.25}
  \pscircle(22,4){2.25}
  \pscircle*(38,4){1}
  \pscircle*(45,4){1}
  \pscircle*(52,4){1}
  \psline(58,4)(68,4)
  \pscircle*[linecolor=white](68,4){2.25}
  \pscircle(68,4){2.25}
  \pscircle*[linecolor=white](88,4){2.25}
  \pscircle(88,4){2.25}
  \psline(86,4)(82,0)
  \psline(86,4)(82,8)
  \psline(68,6)(84,6)
  \psline(68,2)(84,2)
  \rput[b](2,9){$\scriptstyle 1$}
  \rput[b](22,9){$\scriptstyle 2$}
  \rput[b](68,9){$\scriptstyle r-1$}
  \rput[b](88,9){$\scriptstyle r$}
\end{pspicture}%
\]
Its automorphism group is trivial, and any automorphism of
$\gothso_{2r+1}(\bbC)$ is an inner automorphism. One can choose the Lie
algebra $\gothso_{2r+1}(\bbC) \cap \gothd_{2r+1}(\bbC)$ as a Cartan
subalgebra of $\gothso_{2r+1}(\bbC)$. Therefore, for an odd $n$ any
finite order automorphism of $\gothso_n(\bbC)$ is conjugated by an inner
automorphism of $\gothso_n(\bbC)$ to an automorphism given by the
relation (\ref{e:45}) with $h \in \rmSO_n(\bbC) \cap \rmD_n(\bbC)$.
Any finite order automorphism of $\rmSO_n(\bbC)$ for an odd $n$ is
conjugated by an inner automorphism of $\rmSO_n(\bbC)$ to an
automorphism given by the relation (\ref{e:46}) with $h \in
\rmSO_n(\bbC) \cap \rmD_n(\bbC)$.

For an even $n=2r$ the rank of the Lie algebra $\gothso_n(\bbC)$ is
equal to $r$ and the corresponding Dynkin diagram is
\[
\psset{unit=1pt, linewidth=.5pt}%
\begin{pspicture}(0,0)(95,44)
  \psline(2,22)(32,22)
  \psline(58,22)(88,22)
  \psline(88,2)(88,42)
  \pscircle*[linecolor=white](2,22){2.25}
  \pscircle(2,22){2.25}
  \pscircle*[linecolor=white](22,22){2.25}
  \pscircle(22,22){2.25}
  \pscircle*(38,22){1}
  \pscircle*(45,22){1}
  \pscircle*(52,22){1}
  \pscircle*[linecolor=white](68,22){2.25}
  \pscircle(68,22){2.25}
  \pscircle*[linecolor=white](88,22){2.25}
  \pscircle(88,22){2.25}
  \pscircle*[linecolor=white](88,2){2.25}
  \pscircle(88,2){2.25}
  \pscircle*[linecolor=white](88,42){2.25}
  \pscircle(88,42){2.25}
  \rput[b](2,27){$\scriptstyle 1$}
  \rput[b](22,27){$\scriptstyle 2$}
  \rput[b](68,27){$\scriptstyle r-3$}
  \rput[lb](93,20){$\scriptstyle r-2$}
  \rput[lb](93,40){$\scriptstyle r-1$}
  \rput[lb](93,0){$\scriptstyle r$}
\end{pspicture}%
\]
For $r \ne 4$ the only nontrivial automorphism of this Dynkin diagram is
given by the equalities
\begin{equation}
\sigma(i) = i, \quad i = 1, \ldots, r-2, \qquad \sigma(r-1) = r, \qquad
\sigma(r) = r-1. \label{e:48}
\end{equation}
When $r = 4$ the group of automorphisms of the Dynkin diagram consists
of all permutations of the vertices with the numbers $1$, $3$ and $4$.
However, the automorphisms of the Lie algebra $\gothso_8(\bbC)$
corresponding to these additional automorphisms of the Dynkin diagram
cannot be lifted up to automorphisms of the Lie group $\rmSO_8(\bbC)$
and we do not consider them here.

Choose the Lie algebra $\gothso_{2r}(\bbC) \cap \gothd_{2r}(\bbC)$ as a
Cartan subalgebra of $\gothso_{2r}(\bbC)$. In this case the Cartan and
Chevalley generators can be chosen as
\begin{align*}
&h_i = e_{ii} - e_{i+1,i+1} + e_{2r-i, 2r-i} - e_{2r-i+1, 2r-i+1}, && i
= 1, \ldots, r-1, \\
&h_r = e_{r-1, r-1} + e_{r,r} - e_{r+1, r+1} - e_{r+2, r+2}, \\
&x_{+i} = e_{i, i+1} - e_{2r-i, 2r-i+1}, && i = 1, \ldots, r-1, \\
&x_{+r} = e_{r-1, r+1} - e_{r, r+2}, \\
&x_{-i} = e_{i+1, i} - e_{2r-i+1, 2r-i}, && i = 1, \ldots, r-1, \\
&x_{-r} = e_{r+1, r-1} - e_{r+2, r}.
\end{align*}
Now one can get convinced that the linear mapping $\Sigma$ defined by
the relation
\[
\Sigma(x) = u \, x \, u^{-1},
\]
where
\begin{equation}
u = \left(\begin{array}{cccc}
I_{r-1} \\
& 0 & 1 \\
& 1 & 0 \\
& & & I_{r-1}
\end{array} \right), \label{e:49}
\end{equation}
is an automorphism of $\gothso_{2r}(\bbC)$ induced by the automorphism
$\sigma$ of the Dynkin diagram of $\gothso_{2r}(\bbC)$ defined by the
equalities (\ref{e:48}).

Thus, for an even $n$ not equal to $8$ a finite order inner automorphism
of the Lie algebra $\gothso_n(\bbC)$ is conjugated by an inner
automorphism of $\gothso_n(\bbC)$ to an automorphism $A$ given by the
equality (\ref{e:45}), where $h$ is an element of $\rmSO_n(\bbC) \cap
\rmD_n(\bbC)$, and a finite order outer automorphism of
$\gothso_n(\bbC)$ is conjugated by an inner automorphism of
$\gothso_n(\bbC)$ to an automorphism $A$ given by the equality
\begin{equation}
A(x) = (h u) \, x \, (h u)^{-1}, \label{e:50}
\end{equation}
where $h$ is an element of $\rmSO_n(\bbC) \cap \rmD_n(\bbC)$ satisfying
the relation $u \, h \, u^{-1} = h$. In the case of $n = 8$ there are
finite order outer automorphisms of $\gothso_8(\bbC)$ different from the
automorphisms defined by the relation (\ref{e:50}) but they cannot be
lifted to automorphisms of $\rmSO_8(\bbC)$.

It is clear now that for an even $n$ a finite order inner automorphism
of the Lie group $\rmSO_n(\bbC)$ is conjugated by an inner automorphism
of $\rmSO_n(\bbC)$ to an automorphism $a$ given by the equality
(\ref{e:46}), where $h$ is an element of $\rmSO_n(\bbC) \cap
\rmD_n(\bbC)$, and a finite order outer automorphism of $\rmSO_n(\bbC)$
is conjugated by an inner automorphism of $\rmSO_n(\bbC)$ to an
automorphism $a$ given by the equality
\[
a(g) = (h u) \, g \, (h u)^{-1},
\]
where $u$ is given by (\ref{e:49}) and $h$ is an element of
$\rmSO_n(\bbC) \cap \rmD_n(\bbC)$ satisfying the relation $u \, h \,
u^{-1} = h$.

\subsection{Complex symplectic groups} \label{a:c4}

Let for an even $n$ the $n \times n$ matrix $B$ be skew-symmetric and
nonsingular. It can be shown that in this case the Lie group
$\rmGL_n^B(\bbC)$ is isomorphic to the Lie group $\rmGL_n^K(\bbC)$. This
group is called the {\em complex symplectic group\/} and denoted
$\rmSp_n(\bbC)$. The Lie group $\rmSp_n(\bbC)$ is connected and simple.
The corresponding Lie algebra $\gothsp_n(\mathbb C)$ is also simple.

The rank of the Lie algebra $\gothsp_n(\bbC)$ is equal to $r=n/2$ and
the corresponding Dynkin diagram is of the form
\[
\psset{unit=1pt, linewidth=.5pt}%
\begin{pspicture}(0,0)(90,13)
  \psline(2,4)(32,4)
  \pscircle*[linecolor=white](2,4){2.25}
  \pscircle(2,4){2.25}
  \pscircle*[linecolor=white](22,4){2.25}
  \pscircle(22,4){2.25}
  \pscircle*(38,4){1}
  \pscircle*(45,4){1}
  \pscircle*(52,4){1}
  \psline(58,4)(68,4)
  \pscircle*[linecolor=white](68,4){2.25}
  \pscircle(68,4){2.25}
  \pscircle*[linecolor=white](88,4){2.25}
  \pscircle(88,4){2.25}
  \psline(70,4)(74,0)
  \psline(70,4)(74,8)
  \psline(72,6)(88,6)
  \psline(72,2)(88,2)
  \rput[b](2,9){$\scriptstyle 1$}
  \rput[b](22,9){$\scriptstyle 2$}
  \rput[b](68,9){$\scriptstyle r-1$}
  \rput[b](88,9){$\scriptstyle r$}
\end{pspicture}%
\]
The automorphism group of this diagram is trivial, and any automorphism
of $\gothsp_n(\bbC)$ is an inner automorphism. One can choose the Lie
algebra $\gothsp_n(\bbC) \cap \gothd_n(\bbC)$ as a Cartan subalgebra of
$\gothsp_n(\bbC)$. Therefore, any finite order automorphism of
$\gothsp_n(\bbC)$ is conjugated by an inner automorphism of
$\gothsp_n(\bbC)$ to an automorphism given by the relation (\ref{e:45})
with $h \in \rmSp_n(\bbC) \cap \rmD_n(\bbC)$. Any finite order
automorphism of $\rmSp_n(\bbC)$ is conjugated by an inner automorphism
of $\rmSp_n(\bbC)$ to an automorphism given by the relation (\ref{e:46})
with $h \in \rmSp_n(\bbC) \cap \rmD_n(\bbC)$.

\end{document}